%% file: arxivversion.tex
\documentclass[screen,acmtog,nonacm]{acmart}

\usepackage{amsmath}
\usepackage{bm}
\usepackage{color,balance,microtype,booktabs}
\usepackage[fleqn,tbtags]{mathtools}
\usepackage{hyperref} 
\usepackage[capitalise]{cleveref} 
\usepackage[detect-weight]{siunitx}
\usepackage{pifont,xspace}
\usepackage{multirow}
\usepackage{multicol}
\usepackage{algorithm}
\usepackage{enumitem}
\usepackage{stfloats}
\usepackage{textcomp}
\usepackage{algpseudocode}

\usepackage[english]{babel}
\usepackage{amsthm}

\algrenewcommand\alglinenumber[1]{\tiny #1:}
\usepackage{amsmath,amssymb}

\graphicspath{{./images/}}

\def \etal {{\emph{et al}.\thinspace}}

\def \eg {{\emph{e.g}.\thinspace}, }

\usepackage{etoolbox}
\newtoggle{arxivmode}
\newcommand{\arxiv}[2]{%
  \iftoggle{arxivmode}{#1}{#2}%
}
\toggletrue{arxivmode} 

\newcommand{\name}{\textsc{GMT}\xspace}

\usepackage{xcolor}
\usepackage{pifont}

\newcommand{\cmark}{\textcolor{green!60!black}{\ding{51}}} 
\newcommand{\xmark}{\textcolor{red!70!black}{\ding{55}}}   

\AtBeginDocument{%
    }

\makeatletter
\@namedef{ver@everyshi.sty}{}
\makeatother

\begin{document}
\title{\name: A Geometric Multigrid Transformer Solver for Microstructure Homogenization}

\author{Yu Xing}
\affiliation{%
  \institution{Shandong University}
  \city{Qingdao}
  \country{China}
}
\email{xing_yu@mail.sdu.edu.cn}

\author{Yang Liu}
\affiliation{%
  \institution{Microsoft Research Asia}
  \city{Beijing}
  \country{China}
}
\email{yangliu@microsoft.com}

\author{Tianyang Xue}
\affiliation{%
  \institution{The University of Hong Kong}
  \city{Hong Kong}
  \country{China}
}
\email{timhsue@gmail.com}

\author{Lin Lu}
\authornote{Corresponding author}
\affiliation{%
  \institution{Shandong University}
  \city{Qingdao}
  \country{China}
}
\email{llu@sdu.edu.cn}

\begin{abstract}
  \input{src/abstract}
\end{abstract}

\input{figures/teaser}
\maketitle

\authorsaddresses{}

\input{src/1-introduction}

\input{src/2-relatedwork}

\input{src/3-background}

\input{src/4-method}
\input{src/5-implementation}

\input{src/6-result}

\input{src/7-application}

\input{src/8-discussion}

\bibliographystyle{ACM-Reference-Format}
\bibliography{src/ref}

\appendix
\input{src/appendix}

\end{document}

%% file: src/abstract.tex
Lattice metamaterials enable lightweight, multifunctional structures, yet homogenization-based evaluation of their effective properties remains computationally expensive. Neural surrogates offer speed but often lack the accuracy and stability required for engineering-grade simulations.
We introduce \name, a \emph{G}eometric \emph{M}ultigrid \emph{T}ransformer -- a neural solver with high numerical fidelity for fast and reliable lattice homogenization.
\name achieves \emph{architectural alignment} with Geometric Multigrid (GMG) by re-restructuring Point Transformer V3 to operate across sparse GMG hierarchies, capturing long-range dependencies and cross-level interactions essential for multigrid convergence.  To enforce physical consistency, \name incorporates physics-aware positional encoding for  strict enforcement of periodicity and predicts both the finest-level solution and multi-level residual corrections. These predictions deliver a \emph{spectrally-aligned initialization}, enabling end-to-end training under physics-informed and solver-aware losses and requiring only a single GMG V-cycle refinement to reach convergence.
This fusion of neural prediction and numerical rigor achieves relative residual errors of $10^{-5}$ with a $160\times$ speedup over state-of-the-art GPU-based solvers at equivalent accuracy -- particularly at high resolutions (\eg $512^3$), where traditional methods become most costly. We validate \name across mechanical and thermal domains, demonstrate robust generalization to unseen geometries and non-periodic settings, and showcase scalability to high resolutions -- enabling real-time design iteration, multi-scale simulations, high-throughput material discovery, and inverse design.

%% file: figures/teaser.tex
\begin{teaserfigure}
  \centering
  \includegraphics[width=\textwidth]{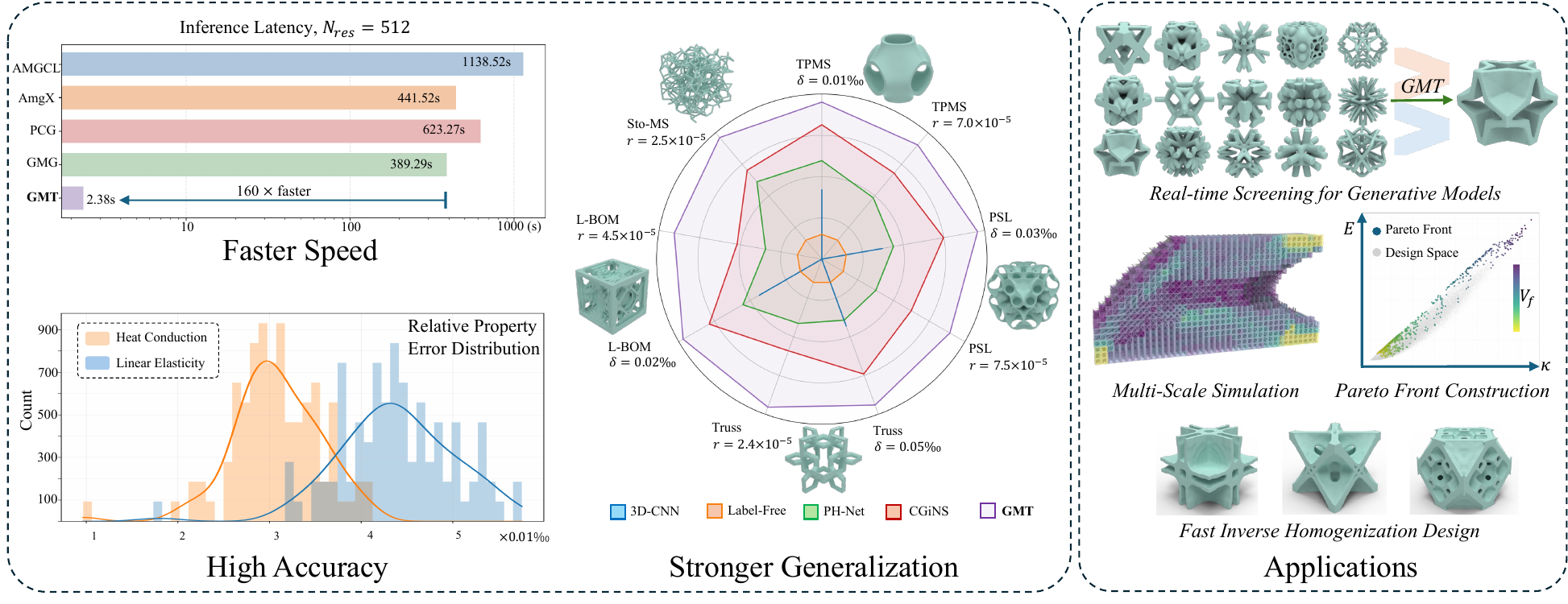}
  \Description{
    We present GMT, a Geometric Multigrid Transformer. It is a high-fidelity, differentiable neural solver for microstructure homogenization.
    On the left, performance: GMT achieves more than a one-hundred-and-sixty-times speedup compared to state-of-the-art GPU solvers such as AmgX or geometric multigrid. This result is obtained at very high resolution, with grids of size five hundred and twelve cubed. Despite this speed, GMT maintains engineering-grade accuracy, with relative errors in predicted material properties below one-tenth of a per-thousand. These results hold across multiple physics problems, including linear elasticity and heat conduction. In the middle, generalization: Unlike previous neural surrogate models, GMT generalizes reliably across many different lattice geometries. These include triply periodic minimal surfaces, plate-surface lattices, stochastic microstructures, lattice-based optimized materials, and truss structures. Importantly, this generalization does not require retraining the model for each geometry type.
    On the right, applications: This computational efficiency enables demanding workflows, such as real-time screening in generative design models, large-scale multiscale simulations, high-throughput inverse design, and construction of Pareto fronts.
  }
  \caption{ We present \name, a Geometric Multigrid Transformer that serves as a high-fidelity, differentiable neural solver for microstructure homogenization. (Left) Superior Performance: \name achieves over $160\times$ speedup compared to state-of-the-art GPU solvers (\eg AmgX or GMG) at high resolutions ($512^3$) while maintaining engineering-grade accuracy (relative property error  $< 0.1$\textperthousand ) across multi-physics domains (linear elasticity and heat conduction). (Middle) Structure Generalization: Unlike previous neural surrogates, \name demonstrates robust generalization across diverse lattice topologies (TPMS, PSL, Sto-MS, L-BOM, Truss) without per-geometry retraining.
    (Right) Applications: This efficiency facilitates computationally demanding workflows, including real-time screening for generative models, large-scale multi-scale simulations, high-throughput inverse design and Pareto front construction.}
  \label{fig:teaser}
\end{teaserfigure}

%% file: src/1-introduction.tex
\section{Introduction}

Microstructures have transformed computational design by embedding functionality into geometry rather than material composition. This paradigm enables lightweight yet high-performance structures across aerospace, mechanical, and thermal management applications \cite{li1996microstructure, schumacher2015microstructures}. Architected lattices deliver exceptional strength, stiffness, and conductivity while offering tunable anisotropy and multifunctional capabilities \cite{kulagin2020architectured, fleck2010micro}.
As additive manufacturing matures, designers can fabricate intricate lattice architectures at sub-millimeter precision using diverse materials \cite{askari2020additive, tan2020microstructure, liu2022additive}. As fabrication becomes increasingly accessible, the bottleneck shifts to physics evaluation: a single design iteration may require thousands of queries of effective properties across geometry variants, loading directions, and coupled-physics settings.
This demand highlights the limitations of existing computational tools \cite{kulagin2020architectured, viswanath2024designing}; while classical solvers offer high  accuracy, they scale poorly with geometric complexity, becoming prohibitive for rapid design iterations.

Recent advances in deep learning offer promising data-driven surrogates that predict homogenized properties \cite{rao2020three} or infer solution fields \cite{silverman2025random, peng2022ph, zhu2024learning}. While fast, these black-box models often struggle on high-contrast geometries and fail to enforce governing physics. More critically, purely neural operators suffer from spectral bias—learning low-frequency modes quickly but struggling with high-frequency details -- leading to drifting errors that preclude high-precision engineering applications.

To overcome these limitations, we propose  \name, a Geometric Multigrid Transformer -- a differentiable neural solver  with high numerical fidelity for fast and reliable lattice homogenization (\cref{fig:teaser}).
\name builds on the principle that \emph{multigrid methods succeed by explicitly decomposing, smoothing, and correcting errors across frequencies}. Unlike loosely coupled hybrid approaches that merely use neural networks as initializers, \name achieves architectural alignment between deep learning and numerical methods through three key designs:
\begin{enumerate}[leftmargin=*]\setlength\itemsep{1mm}
    \item \emph{GMG-aligned architecture}: by re-architecting Point Transformer V3~\cite{wu2024point} to operate directly on sparse Geometric Multigrid (GMG) hierarchies, enabling the network to capture long-range dependencies and cross-level interactions fundamental to multigrid solvers.
    \item \emph{Physics-aware encoding}: introducing resolution-aware rotary relative positional encoding (Ra-RoPE) to handle strict periodic boundary conditions (PBCs) and lattice symmetries.
    \item \emph{Spectrally-aligned initialization}: predicting both the finest-level solution and multi-level residuals, to deliver a spectrally aligned initialization, enabling convergence within a single V-cycle.
\end{enumerate}

The entire system is trained end-to-end with physics-informed and GMG-solver-aware losses, complemented by a GPU-oriented element-by-element GMG implementation, enabling rapid numerical correction while preserving gradient flow for downstream optimization tasks.

Our deep fusion of neural prediction with numerical rigor delivers a substantial improvement in performance: \emph{\name achieves relative residual errors of $10^{-5}$ -- a precision unattainable by purely neural surrogates -- while offering over $160\times$ speedup compared to state-of-the-art GPU solvers at equivalent accuracy -- particularly at high resolutions \eg $512^3$}. We validate \name across mechanical and thermal domains, demonstrate robust generalization to unseen geometries and non-periodic settings, and showcase scalability to high-resolution, enabling real-time design iteration, high-throughput material discovery, and inverse design applications. We will release our code and model at \url{https://github.com/xing-yuu/GMT} to facilitate future research.

%% file: src/2-relatedwork.tex
\section{Related Work}
\label{sec:relatedwork}

\subsection{Homogenization Computation}

\paragraph{Numerical Homogenization} Homogenization theory provides a rigorous PDE framework for linking microscale geometry to macro-scale performance by solving boundary value problems on Representative Volume Elements (RVEs)~\cite{michel1999effective}.
This paradigm has been instrumental in modeling material appearance and physical behavior across scales. Foundational work by~\cite{schumacher2015microstructures} leveraged tileable microstructures to achieve target elastic properties in 3D printing. More recently, homogenization has been extended to complex thin-shell structures via polar interpolants~\cite{chan2024polar} and to the discrete mechanics of textiles, where~\cite{sperl2020homogenized} developed homogenized models for yarn-level cloth.
Several numerical strategies have been proposed, including spectral and FFT-based schemes~\cite{schneider2021review, chen2020comparison}, boundary element methods~\cite{prochazka2003bem}, and mean-field approximations~\cite{sekkate2022elastoplastic}. Among these, the Finite Element Method (FEM) remains the most widely adopted due to its versatility across complex geometries and multiphysics domains~\cite{dong2019149, yvonnet2019computational}. However, FEM-based homogenization requires solving large, sparse linear systems derived from fine voxelizations, imposing a severe computational burden~\cite{hughes2003finite}. Iterative solvers such as Conjugate Gradient (CG) with Incomplete Cholesky preconditioning, and multilevel approaches like Algebraic Multigrid (AMG)~\cite{ruge1987algebraic} and Geometric Multigrid (GMG)~\cite{clevenger2020flexible, sampath2010parallel}, have improved scalability. GPU acceleration further enhances throughput~\cite{cecka2011assembly, dick2011real, haase2010parallel}. Nevertheless, achieving high accuracy for complex, high-contrast lattices still demands numerous iterations~\cite{xu2008uniform}, making these methods impractical for interactive design loops or simulations involving millions of unit cells~\cite{geers2010multi}. We retain multigrid rigor but reduce computational cost by requiring only a single V-cycle after spectrally-aligned neural initialization.

\paragraph{Neural Homogenization} Early data-driven methods focused on surrogate modeling, using convolutional neural networks (CNNs) to map geometry directly to homogenized properties~\cite{rao2020three}.
Beyond linear regimes, Li \etal~\shortcite{li2023neural} introduced neural metamaterial networks for nonlinear homogenization and optimization over parametric 2D structures. For cloth simulation, Feng \etal~\shortcite{feng2024neural} presented neural-assisted homogenization to bridge the gap between yarn-level detail and efficient macroscale simulation.
Subsequent works advanced this paradigm by predicting full solution fields~\cite{peng2022ph, zhu2024learning, silverman2025random}, enabling instant feedback but suffering from poor generalization and catastrophic failures outside the training distribution. Recent work has explored spectral alignment strategies, such as differentiable PCG solvers that emphasize high projection energy modes~\cite{xing2025pcg}. While effective in reducing low-frequency bias, these methods remain constrained by the convergence properties of single-level Krylov solvers, fundamentally limiting achievable accuracy and scalability.
In contrast, we advance from spectral alignment to \textit{architectural alignment}, embedding the neural predictor directly within a GMG hierarchy to ensure full-spectrum error correction.

\subsection{General Neural PDE Solvers}
\paragraph{End-to-End Surrogates}
Physics-Informed Neural Networks~\cite{raissi2019physics, cai2021physics} enforce PDE constraints via soft loss terms, while Neural Operators such as Fourier Neural Operator (FNO)~\cite{li2020fourier} and DeepONet~\cite{lu2021learning} learn mappings between infinite-dimensional function spaces.
In the context of structural design, Zehnder \etal~\shortcite{zehnder2021ntopo} explored the use of implicit neural representations (NTopo) for mesh-free topology optimization.
Recent work has also explored Large Language Models for PDE analysis~\cite{zhou2024unisolver}. While promising, these methods are primarily validated on simple PDEs (\eg Poisson) and often suffer from spectral bias, struggling to capture high-frequency components critical for engineering-grade accuracy. Operator-learning frameworks also require massive labeled datasets~\cite{brandstetter2022message}, which are computationally expensive to generalize to complex 3D physics problems.  Our approach avoids spectral bias by coupling neural prediction with the multigrid hierarchy and ensures solver-level accuracy.

\paragraph{Hybrid Approaches}
Neural-augmented solvers accelerate intermediate steps~\cite{luz2020learning}, learn preconditioners~\cite{trifonov2024learning, azulay2022multigrid, chen2024graph, lan2024neural}, or provide initial guesses~\cite{um2020solver, pestourie2023physics}. However, most hybrids adopt a loosely coupled paradigm, treating the numerical solver as a black-box post-processor. Critically, these solvers lack hierarchical structures to decompose error frequencies effectively. Even with high-quality neural initialization, single-level solvers struggle to eliminate residuals across the entire spectrum, particularly for multi-scale microstructures.  We integrate the neural model into the GMG hierarchy, aligning prediction with solver correction for efficient, frequency-aware convergence.

%% file: src/3-background.tex
\section{Background}

\subsection{Periodic Microstructure Homogenization}
\label{sec:problem_statement}

Homogenization is formulated on a periodic representative volume element (RVE) $\Omega \subset \mathbb{R}^3$, represented by a binary voxel field representing fixed constituent materials.
Let $\boldsymbol{\Upsilon} \in \mathbb{R}^3$ denote the period vector (the lattice cell size along each axis), and let $N_{\mathrm{res}}$ denote the voxel resolution along each dimension, so that opposite faces of $\Omega$ are discretized with $N_{\mathrm{res}}$ voxels along each axis.
The objective is to compute homogenized tensors mapping macroscopic loading fields to volume-averaged microscopic responses.

A broad class of homogenization problems reduces to periodic cell problems governed by linear elliptic PDEs. For each macroscopic loading mode $m = 1,\dots,M$, the periodic corrector $\mathbf{u}^{(m)}$ satisfies:
\begin{equation}
    \nabla \cdot \bigl( \boldsymbol{\Lambda}(\mathbf{x})(\nabla \mathbf{u}^{(m)} + \mathbf{g}^{(m)}) \bigr) = 0 \quad \text{in } \Omega,
    \label{eq:generic_cell}
\end{equation}
subject to periodic boundary conditions:
\begin{equation}
        \mathbf{u}^{(m)}(\mathbf{x})= \mathbf{u}^{(m)}(\mathbf{x} \pm \boldsymbol{\Upsilon}),
        \quad  \mathbf{x} \in \partial\Omega,
    \label{eq:pbc_generic}
\end{equation}
together with a Gauge constraint (\eg $\langle \mathbf{u}^{(m)}\rangle_\Omega = 0$, where $\langle \cdot \rangle_\Omega$ denotes the volume average) for uniqueness. Here $\boldsymbol{\Lambda}$ denotes the material coefficients (\eg stiffness or conductivity). The term $\mathbf{g}^{(m)}$ encodes the imposed macroscopic mode. Discretization of \cref{eq:generic_cell} on a fine voxel grid yields a large sparse linear system:
\begin{equation}
    \mathbf{K}(\Omega)\,\mathbf{u} = \mathbf{f}(\Omega),
    \label{eq:linear_system}
\end{equation}
with millions of degrees of freedom. Thin struts, high-contrast materials, and near-disconnected geometries make the resulting systems challenging for iterative solvers, slowing convergence and increasing the number of iterations required.

We consider two representative physics problems: linear elasticity (as a vector-valued PDE) and steady heat conduction (as a scalar PDE).

\paragraph{Example 1: Linear Elasticity (Effective Stiffness)}
The homogenized elasticity tensor $\mathbf{C}_H \in \mathbb{R}^{6\times 6}$ is obtained by solving six cell problems for independent unit strain modes. For each mode $kl$ ($k,l\in\{x,y,z\}$), the displacement corrector $\boldsymbol{\chi}_{kl}$ satisfies:
\begin{equation}
        \begin{aligned}
             & \nabla \cdot \bigl( \mathbf{C}_0 \colon (\boldsymbol{\epsilon}(\boldsymbol{\chi}_{kl}) + \boldsymbol{\epsilon}^{0}_{kl}) \bigr) = 0 \quad \text{in } \Omega, \\
             & \boldsymbol{\chi}_{kl}(\mathbf{x})=\boldsymbol{\chi}_{kl}(\mathbf{x}\pm\boldsymbol{\Upsilon}), \quad \forall \mathbf{x}\in\partial\Omega.
        \end{aligned}
    \label{eq:elastic_cell}
\end{equation}

The operator $:$ means the double dot product of two tensors. The microscopic stress is $\boldsymbol{\sigma}_{kl} = \mathbf{C}_0 \colon (\boldsymbol{\epsilon}(\boldsymbol{\chi}_{kl})+\boldsymbol{\epsilon}_{0,kl})$, where $\mathbf{C}_0$ is the base elasticity tensor, and $\mathbf{C}_H$ follows by volume averaging.

\paragraph{Example 2: Steady Heat Conduction (Effective Thermal Conductivity)}
The homogenized thermal conductivity $\boldsymbol{\kappa}_H \in \mathbb{R}^{3\times 3}$ is computed by solving three scalar cell problems for unit temperature gradients. For each direction $i\in\{x,y,z\}$:
\begin{equation}
    \begin{aligned}
         & \nabla \cdot \big( \kappa_0(\mathbf{x}):(\nabla \psi_i + \mathbf{e}_i) \big) = 0 \quad \text{in } \Omega, \\
         & \psi_i(\mathbf{x})=\psi_i(\mathbf{x}+\boldsymbol{\Upsilon}), \quad \forall \mathbf{x}\in\partial\Omega.
    \end{aligned}
    \label{eq:thermal_cell}
\end{equation}
The microscopic heat flux is $\mathbf{q}_i = -\kappa_0(\nabla \psi_i + \mathbf{e}_i)$, and $\boldsymbol{\kappa}_H$ is obtained by volume averaging.

More details about homogenization are provided in \arxiv{\cref{appendix:F}}{Suppl.~Sec.~6}.

\subsection{Geometric Multigrid (GMG) Framework}
\label{sec:gmg_framework}

The Geometric Multigrid (GMG) method achieves optimal $O(N)$ complexity for solving large-scale linear systems by decomposing error components across a hierarchy of grids~\cite{dick2011real}.

\paragraph{Grid Hierarchy}
A sequence of $L$ nested grids $\{\Omega_l\}_{l=1}^L$ is constructed, where $l=1$ denotes the finest resolution and $l=L$ the coarsest.
The hierarchy is defined by a progressive grid coarsening. As the density of active degrees of freedom decreases, spatially smooth errors on the fine grid appear oscillatory  under the coarser discretization, enabling efficient correction.

\paragraph{V-Cycle Algorithm}
A standard GMG V-cycle recursively eliminates error modes across the frequency spectrum. It alternates between damping high-frequency errors via Gauss-Seidel ($\mathrm{GS}$) smoothing and resolving low-frequency residuals on coarser grids (see \cref{alg:vcycle}). Multiple cycles are typically required to reach the target tolerance.

\begin{algorithm}[t]
    \begin{footnotesize}
        \caption{Standard GMG V-Cycle}
        \label{alg:vcycle}
        \begin{algorithmic}[1]
            \State ${\mathbf{u}}^1 \leftarrow 0$ \Comment{Initialize solution}
            \For{$l = 1, \dots, L-1$}
            \State ${\mathbf{u}}^l \leftarrow \mathrm{GS}(\mathbf{K}^l, \mathbf{u}^l, \mathbf{f}^l; \mathbf{It}^l)$ \Comment{Pre-Smoothing}
            \State $\mathbf{r}^l \leftarrow \mathbf{f}^l - \mathbf{K}^l \mathbf{u}^l$ \Comment{Residual update}
            \State $\mathbf{f}^{l+1} \leftarrow \mathbf{R} \mathbf{r}^l$
            \State ${\mathbf{u}}^{l+1}\leftarrow 0$ \Comment{Initialize coarse error}
            \EndFor
            \State ${\mathbf{u}}^L \leftarrow \mathrm{GS}(\mathbf{K}^L, \mathbf{u}^L, \mathbf{f}^L; \mathbf{It}^L)$ \Comment{Coarsest-level solve}

            \For{$l = L-1, \dots, 1$}
            \State $\mathbf{u}^l \leftarrow \mathbf{u}^l + \mathbf{P} \mathbf{u}^{l+1}$ \Comment{Prolongate and correct}
            \State ${\mathbf{u}}^l \leftarrow \mathrm{GS}(\mathbf{K}^l, \mathbf{u}^l, \mathbf{f}^l; \mathbf{It}^l)$ \Comment{Post-Smoothing}
            \EndFor
            \State \Return $\mathbf{u}^1$
        \end{algorithmic}
    \end{footnotesize}
\end{algorithm}

\paragraph{Operator Consistency}
Coarse-level operators are constructed via Galerkin projection:
\[
    \mathbf{K}^{(l+1)} = \mathbf{R}\,\mathbf{K}^{(l)}\,\mathbf{P},
\]
where $\mathbf{P}$ denotes the prolongation (interpolation) operator and $\mathbf{R}$ the restriction operator, typically $\mathbf{R} = \mathbf{P}^\mathsf{T}$.  This hierarchy forms the foundation for efficient GMG solvers and underpins the design choices in our \name.

%% file: src/4-method.tex
\section{Method}
\label{sec:method}

\paragraph{Overview}
\name introduces architectural alignment to deeply couple neural prediction with the GMG solver. This design exploits spectral complementarity: the neural network resolves global low-frequency dependencies, while the numerical backend efficiently eliminates local high-frequency errors. To realize this aligned architecture under strict physical constraints and ensure high-throughput execution, our framework is grounded in six core principles:

\begin{enumerate}[leftmargin=*]
    \item \textbf{Sparse GMG Hierarchy:}  Build from active geometry to reduce memory overhead while preserving multilevel error decomposition (\cref{sec:sparse_gmg_hierarchy}).
    \item \textbf{GMG-Aligned Neural Architecture:} Align transformer-based design with GMG levels to efficiently capture long-range dependencies and cross-level interactions (\cref{sec:neural_architecture}).
    \item \textbf{Physics-Aware Encoding:}  Encode strict PBCs by Resolution-Aware Rotary Positional Encoding (Ra-RoPE) within attention (\cref{sec:ra_pe}).
    \item \textbf{Spectrally-Aligned Initialization:} Predict both finest-level solution and multi-level residuals jointly to warm-start the GMG V-cycle and accelerate convergence to engineering-grade accuracy (\cref{sec:spectral_initialization}).
    \item \textbf{Solver-Aware Loss Design:} Train with physics-informed and GMG-aware objectives in an end-to-end manner to ensure numerical consistency (\cref{sec:physical_loss}).
    \item \textbf{Element-by-Element V-Cycle:} Implement matrix-free GMG V-cycle to enable fine-grained GPU parallelism (\cref{sec:ebe_gmg}).
\end{enumerate}
Together, these components combine neural efficiency with multigrid rigor under strict physical constraints, delivering convergence in a single V-cycle for engineering-grade accuracy. \cref{fig:pipeline} illustrates the overall architecture of \name.
\input{figures/pipeline}


\subsection{Sparse GMG Hierarchy}
\label{sec:sparse_gmg_hierarchy}

\subsubsection{Domain Discretization}
\label{subsubsec:Discretization}
The homogenization problem is discretized on a high-resolution voxel grid within a periodic RVE, consisting of $n_e = N_{\text{res}}^3$ elements. To reduce memory overhead, \name operates on \emph{sparse voxels} rather than the full dense grid.

Formally, the network input is the set of active nodes $\mathcal{P}_1 = \{ \mathbf{p}_i \}_{i=1}^{N_{\text{active}}}$, where each node $ \mathbf{p}_i \in [0,\boldsymbol{\Upsilon})^3$ corresponds to a voxel vertex in the periodic  material domain. Each node carries an 8-dimensional binary feature vector $\mathbf{f}_i^1 \in \{0,1\}^8$ encoding the occupancy of the eight surrounding octants (voxels) sharing that vertex. This representation captures local topology and material connectivity while excluding void regions, enabling scalability to high resolutions such as $512^3$ that would otherwise be memory-prohibitive. We focus primarily on the impact of structural variations on effective properties; thus, the material parameters ($\boldsymbol{\Lambda}(\mathbf{x})$) are set as constants. This does not imply that our framework is limited to uniform materials, as its extension to heterogeneous and high-contrast settings is detailed in \arxiv{\cref{appendix:D1}}{Suppl.~Sec.~4.1}.

\subsubsection{Solver-Aligned Hierarchy Construction}
Unlike solver-agnostic coarsening in generic point-cloud and sparse-voxel backbones, such as PTv3~\cite{wu2024point}, which coarsens points via grid pooling, or MinkowskiNet~\cite{choy20194d,choy2020high}, which voxelizes the input and downsamples features through stride-2 sparse convolutions, our hierarchy is \emph{deterministically aligned} with the GMG solver topology.

To strictly mirror numerical restriction, we define voxel activity recursively following standard hexahedral GMG~\cite{dick2011real}: a coarse voxel is included in the active set $\mathcal{E}_{l+1}$ if it spatially encompasses \emph{any} active fine voxel from $\mathcal{E}_l$. This conservative rule preserves topological connectivity for thin features. The set of active nodes $\mathcal{P}_{l+1}$ is then derived as the vertex union of $\mathcal{E}_{l+1}$:
\begin{equation}
    \mathcal{P}_{l+1} = \{ x \in \mathbb{Z}^3 \mid \exists e \in \mathcal{E}_{l+1} \text{ s.t. } x \in \text{vertices}(e) \}.
    \label{eq:gmgcoarsenode}
\end{equation}
By aligning neural receptive fields with the mathematical restriction stencil, the network effectively internalizes the multiscale interaction logic fundamental to multigrid theory.


\subsection{GMG-Aligned Neural Network Structure} \label{sec:neural_architecture}

\subsubsection{Network Architecture}
The strong variable coupling observed in Gauss-Seidel iterations in GMG solvers motivates the need for modeling long-range dependencies in the neural counterpart. In GMG, Gauss-Seidel updates each node sequentially, where the value of one node strongly depends on its neighbors and, through residual propagation, on distant nodes across the grid. This iterative dependency means that local updates influence global error modes, especially in high-contrast lattices where stress or flux propagates over long distances. Capturing these interactions efficiently requires a mechanism that aggregates both local and global context.

To address this, \name builds upon Point Transformer V3 (PTv3)~\cite{wu2024point}, which in its original design processes sparse point clouds using serialization-based window attention. Nodes are mapped into a 1D sequence via space-filling curves (\eg Morton/Z-order), enabling efficient attention with linear complexity. We restructure PTv3 to incorporate a sparse GMG hierarchy to align it with the GMG solver structure and homogenization physics.

\paragraph{Overall Architecture}
\name adopts a symmetric U-Net-like design aligned with the hierarchical levels of a GMG V-cycle (see~\cref{fig:pipeline}, left). The network consists of a down-sampling path and an upsampling path connected by skip connections across $L$ grid levels.

\paragraph{Downsampling Path (Restriction Phase):} This path parallels the restriction phase of the V-cycle. Each level applies Homogenization-Aware Serialization blocks (\cref{subsubsec:serialization}) to capture geometric features, followed by GMG-Aware Pooling (\cref{subsubsec:gmg_pooling}) to transfer features to the next coarser grid $\Omega_{l+1}$. This process mirrors numerical residual restriction.

\paragraph{Upsampling Path (Prolongation Phase):} This path mirrors the prolongation and correction phase. Features from coarser levels are expanded via symmetric GMG-Aware Unpooling (\cref{subsubsec:gmg_pooling}) and fused with fine-scale features through skip connections. This ensures that long-range interactions captured at coarse scales propagate effectively back to the fine grid.

\paragraph{Network input}
\name takes the sparse active node set $\mathcal{P}_1$ (defined in~\cref{subsubsec:Discretization}) as input. To transform the discrete geometry into a continuous representation, we lift the 8-dimensional occupancy descriptors $f_i^1$ into a $d$-dimensional feature space using a learnable embedding layer: $h_i^1 = \phi_{in}(f_i^1) \in \mathbb{R}^d$. These initial features are then processed through the hierarchical backbone, where periodic spatial relations are encoded via Homogenization-aware serialization (\cref{subsubsec:serialization}) and Resolution-Aware Rotary Positional Encoding (\cref{sec:ra_pe}).

\paragraph{Network output}
Unlike traditional surrogates that only predict a single-scale solution, \name produces hierarchical predictions aligned with the GMG hierarchy $\{\mathcal{P}_l\}_{l=1}^L$. A lightweight task-specific head $\zeta_l$ maps backbone features $h_i^l$ to the physical space at each level, but with distinct objectives: at the finest level ($l=1$), it predicts the full solution field $\hat{u}^1$; at coarser levels ($l > 1$), it predicts the error corrections $\hat{e}^l$.
The output dimensions are determined by the physics: $\mathbb{R}^{3 \times 6}$ for linear elasticity (3D displacements under 6 strain modes) and $\mathbb{R}^{1 \times 3}$ for steady heat conduction (scalar temperatures under 3 gradient modes).
Collectively, these predictions provide a spectrally-aligned initialization, where $\hat{u}^1$ serves as the global guess and $\{\hat{e}^l\}_{l=2}^L$ are injected as coarse-grid corrections to warm-start the subsequent GMG V-cycle (\cref{sec:spectral_initialization}).


\subsubsection{Homogenization-Aware Serialization}
\label{subsubsec:serialization}
\input{figures/Isotropic-SA}

Window attention based on a single serialization order suffers from inherent spatial discontinuities, where physically adjacent voxels often map to distant 1D positions, creating anisotropic blind spots in the receptive field. Although generic backbones like PTv3 mitigate this by mixing heterogeneous patterns (\eg Z-order and Hilbert), such stochastic strategies are suboptimal for lattice-based homogenization. Specifically, mixing curves with varying locality properties fails to guarantee the isotropic aggregation required by physically symmetric operators, and the resulting inconsistent neighborhood definitions across layers disrupt the strict periodicity essential for accurate homogenization.

To address these issues, we propose \emph{Homogenization-Aware Serialization} tailored to voxel lattices (see~\cref{fig:IsotropicSA}). Instead of mixing curve families, we construct three complementary Morton views under cyclic coordinate permutations: $(x,y,z)$, $(y,z,x)$, and $(z,x,y)$. These views are shuffled across attention layers and fed to window attention. This structured set of traversals systematically bridges adjacency gaps, ensuring isotropic coverage with minimal complexity. By aggregating contexts from these three views, the network achieves a comprehensive receptive field that is robust to lattice orientation and consistent with the physical symmetry of homogenization operators. Empirical results in~\cref{tab:ablation_components} confirm the effectiveness of this strategy.

\subsubsection{GMG-aware Pooling and Unpooling} \label{subsubsec:gmg_pooling}
\input{figures/pooling}

To maintain strict alignment with the GMG solver, we introduce \emph{GMG-Aware Pooling} built on the deterministic hierarchy defined in \cref{eq:gmgcoarsenode}. Pooling from level $l$ to $l+1$ operates only on solver-valid coarse nodes $x \in \mathcal{P}_{l+1}$, each anchored to its fine-grid counterpart $2x \in \mathcal{P}_{l}$ for index mapping and stencil construction. This ensures that the neural down-sampling path mirrors the restriction phase of the GMG V-cycle.

Feature aggregation -- the neural counterpart of numerical restriction -- uses a $3^3$ nodal stencil centered at $2x $ and learns a data-adaptive restriction via an MLP:
\begin{equation}
    \mathcal{F}_{l+1}(x) =
    \text{MLP}_{\text{pool}}
    \Big(
    \text{Concat}\big(\{\mathcal{F}_l(2x+k)\mid k\in\{-1,0,1\}^3\}\big)
    \Big).
\end{equation}

To achieve structural isomorphism, we implement a symmetric unpooling path that mirrors numerical prolongation. Unpooling uses the same connectivity and index mapping as pooling, with a corresponding MLP for feature expansion:
\begin{equation}
    \mathcal{F}_{l}^{\text{up}}(2x+k)=\text{MLP}_{\text{up}}(\mathcal{F}_{l+1}(x)).
\end{equation}

This design guarantees that feature transfers across scales follow the same topology as GMG restriction and prolongation, enabling the network to internalize multigrid-style residual propagation and error correction.
\cref{fig:pooling} illustrates our pooling strategy alongside common voxel-based pooling. The ablation study (\cref{subsec:ablation}) confirms the effectiveness of this strategy for homogenization accuracy.


\subsection{Physics-Aware Encoding for Periodicity}
\label{sec:ra_pe}

Homogenization requires strict enforcement of periodic boundary conditions (PBCs), as defined in~\cref{eq:pbc_generic}. Standard positional encodings treat nodes at $p_i = 0$ and $p_i = \boldsymbol{\Upsilon}_i$ as distant, breaking physical continuity across periodic interfaces. Existing approaches often rely on costly periodic convolutions~\cite{xing2025pcg} or soft loss-based constraints~\cite{peng2022ph,zhu2024learning}, which are either computationally expensive or approximate.

To address this, we introduce \emph{Resolution-Aware Rotary Relative Positional Encoding (Ra-RoPE)}, which enforces periodicity natively within self-attention. Unlike standard RoPE~\cite{su2024roformer}, which assumes fixed angular frequencies tied to token positions, Ra-RoPE parameterizes rotations using the physical period $\boldsymbol{\Upsilon}_i$ of the unit cell. This guarantees phase continuity across periodic boundaries and scales naturally with GMG levels.

\paragraph{Formulation}
For a node at coordinate $\mathbf{p}_n = (p_{n,1}, \ldots, p_{n,D})$ within a periodic domain of size $\boldsymbol{\Upsilon}$, Ra-RoPE applies rotations to the query and key vectors. We decompose the feature channels into subspaces for each spatial axis. For a specific axis $i$, let $d_{sub}$ be the allocated dimension. We group these channels into $d_{sub}/2$ pairs, indexed by $k \in \{0, \dots, d_{sub}/2 - 1\}$.

The rotation angle $\theta_{n,i,k}$ for the $k$-th channel pair is parameterized as a harmonic series:
\begin{equation}
    \theta_{n,i,k} = \frac{2\pi (k+1)}{\boldsymbol{\Upsilon}_i} \, p_{n,i}.
    \label{eq:rarope_theta}
\end{equation}
Channels are split across attention heads and axes, and each $(q^{(r)}, q^{(i)})$ pair is rotated by $\theta_{n,i,k}$ using a standard $2\times2$ rotation matrix.

\paragraph{Periodicity Guarantee}
For any integer $t$, shifting by $t\boldsymbol{\Upsilon}_i$ leaves the rotation unchanged:
\begin{equation}
    \theta(p_{n,i} + t\boldsymbol{\Upsilon}_i) - \theta(p_{n,i}) = \frac{2\pi(k+1)}{\boldsymbol{\Upsilon}_i} \cdot t\boldsymbol{\Upsilon}_i = 2\pi(k+1)t \in 2\pi\mathbb{Z}.
\end{equation}
Thus, nodes at $p_i = 0$ and $p_i = \boldsymbol{\Upsilon}_i$ produce identical embeddings, enforcing strict PBCs across all GMG levels.

We apply Ra-RoPE to input features and all attention layers in our implementation.

\subsection{Spectrally-Aligned Initialization}
\label{sec:spectral_initialization}
Unlike existing neural approaches that treat field prediction as a single-level inference task, \name functions as a hierarchical, frequency-aware neural corrector. By explicitly aligning the network's feature hierarchy with the GMG grid levels $\{\mathcal{P}_l\}_{l=1}^L$, \name functions not just as a black-box surrogate, but as a structured component that eliminates error modes across the entire spectrum.

\paragraph{Multi-level Prediction Heads}
To facilitate this, we attach a lightweight, level-specific prediction head $\zeta_l$ to the upsampling path of the backbone at each level $l$. Each head consists of a three-layer convolutional module that maps the $d$-dimensional nodal features $h_i^l$ to the physical solution space:
\begin{enumerate}[leftmargin=*]\setlength\itemsep{1mm}
    \item \emph{Finest-level Head ($l=1$)}: Predicts the full solution field $\hat{u}^1$. This prediction captures high-frequency geometric details and the primary response to the macroscopic loading, serving as the global initial guess.
    \item \emph{Coarse-level Heads ($l > 1$)}: Instead of full solutions, these heads are trained to predict the hierarchical \emph{corrections} $\hat{e}^l$. This design allows the network to internalize the multigrid error correction logic, predicting the adjustment required at each coarser scale to resolve low-frequency residual errors.
\end{enumerate}
\paragraph{Spectral Alignment and Warm-start.}
The output dimensions of $\zeta_l$ are strictly determined by the physics domain: $18$ channels ($\mathbb{R}^{3 \times 6}$) for linear elasticity and $3$ channels ($\mathbb{R}^{1 \times 3}$) for steady heat conduction. As illustrated in the EBE-GMG solver (\cref{fig:pipeline}, right), these hierarchical predictions are explicitly injected into the numerical solver to bypass the typical warm-up phase. Specifically, following~\cref{alg:gmt_vcycle}, the finest-level prediction $\hat{u}^1$ initializes the solution field (Line 1), while the coarse-level predictions $\hat{e}^l$ are injected as initial error corrections during the restriction phase (Line 6). By providing this spectrally-aligned initialization, \name effectively resolves the global, long-range dependencies that typically stall iterative solvers, leaving the numerical smoothers to efficiently handle only the remaining local high-frequency errors. This deep fusion ensures that the numerical solver starts from a high-fidelity state, enabling terminal convergence to engineering-grade accuracy ($10^{-5}$ relative residual) within a single V-cycle.

\begin{algorithm}[t]
    \begin{footnotesize}
        \caption{GMG V-Cycle in \name}
        \label{alg:gmt_vcycle}
        \begin{algorithmic}[1]
            \Require Network outputs: $\textcolor{brown}{\hat{\mathbf{u}}^{1}}$ and $\textcolor{brown}{\{\hat{\mathbf{e}}^{l}\}_{l=2}^{L}}$
            \State $\mathbf{u}^1 \leftarrow \textcolor{brown}{\hat{\mathbf{u}}^{1}}$ \Comment{Spectrally-aligned initialization from network}
            \For{$l = 1, \dots, L-1$}
            \State ${\mathbf{u}}^l \leftarrow \mathrm{GS}(\mathbf{K}^l, \mathbf{u}^l, \mathbf{f}^l; \mathbf{It}^l)$ \Comment{Pre-Smoothing}
            \State $\mathbf{r}^l \leftarrow \mathbf{f}^l - \mathbf{K}^l \mathbf{u}^l$ \Comment{Residual update}
            \State $\mathbf{f}^{l+1} \leftarrow \mathbf{R} \mathbf{r}^l$ \Comment{Restrict residual}
            \State $\mathbf{u}^{l+1} \leftarrow \textcolor{brown}{\hat{\mathbf{e}}^{\,l+1}}$ \Comment{Inject network-predicted correction}
            \EndFor

            \State ${\mathbf{u}}^L \leftarrow \mathrm{GS}(\mathbf{K}^L, \mathbf{u}^L, \mathbf{f}^L; \mathbf{It}^L)$ \Comment{Coarsest-level solve}

            \For{$l = L-1, \dots, 1$}
            \State $\mathbf{u}^l \leftarrow \mathbf{u}^l + \mathbf{P} \mathbf{u}^{l+1}$ \Comment{Prolongate and correct}
            \State ${\mathbf{u}}^l \leftarrow \mathrm{GS}(\mathbf{K}^l, \mathbf{u}^l, \mathbf{f}^l; \mathbf{It}^l)$ \Comment{Post-Smoothing}
            \EndFor
            \State \Return $\mathbf{u}^1$
        \end{algorithmic}
    \end{footnotesize}
\end{algorithm}

\subsection{Physical Loss Design}
\label{sec:physical_loss}

Unlike conventional surrogate models that rely on supervised learning with expensive FEM labels, \name is trained in a \emph{label-free}, physics-informed manner.  The objective is to minimize violations of the governing discrete PDE system.

\paragraph{Residual-Based Loss and Gauge Constraint}
The primary loss is defined directly as the $L^2$-norm of the residual (\cref{eq:linear_system}):
\begin{equation}
    \mathcal{L}_{\text{res}} = \| \mathbf{f} - \mathbf{K}\mathbf{u}^1 \|_2,
\end{equation}
where $\mathbf{u}^1$ denotes the final solution field output by~\cref{alg:gmt_vcycle}.

To ensure uniqueness for periodic cell problems, we impose  a zero-mean displacement constraint: $\sum_{i=1}^{N_{\text{active}}} u^1_i = 0$ (analogous to imposing a Dirichlet boundary condition for eliminating rigid body modes).
This global constraint replaces traditional point-fixing, avoiding artificial stiffness and sharp gradients that hinder convergence. It provides a smoother optimization landscape and preserves translation invariance.

\paragraph{Numerical Stability and Log-Loss}
At high precision, the standard $L^2$ loss becomes unstable when residuals fall below $10^{-3}$, as the loss scale contracts rapidly, causing gradient magnitudes to diminish and making optimization progress increasingly slow. To address this, we apply a logarithmic transformation:
\begin{equation}
    \mathcal{L}_{\text{log}} = \log_{10}\big( \| \mathbf{f} - \mathbf{K}\mathbf{u}^1 \|_2 + 10^{-12}\big).
\end{equation}
The log loss maintains consistent gradient scaling across multiple orders of magnitude, effectively stretching the optimization landscape at low residual levels. This enables \name to achieve stable convergence down to a relative residual of $10^{-5}$ -- a precision previously unattainable by pure neural surrogates.

The total loss combines the log-residual term with a Gauge penalty:
\begin{equation} \label{eq:total_loss}
    \mathcal{L}_{\text{total}} = \mathcal{L}_{\text{log}} + \lambda \left( \sum_{i=1}^{N_{\text{active}}} u_i^1 \right)^2,
\end{equation}
where $\lambda = 10^{-2} \times\eta$ is a hyperparameter proportional to the learning rate $\eta$, enforcing a gradually relaxing Gauge constraint throughout the training process.

\subsection{Element-by-Element GMG Framework}
\label{sec:ebe_gmg}

While the neural component of \name provides a spectrally-aligned initialization, a robust numerical backbone is essential for terminal convergence and physical rigor. We implement a GPU-optimized and matrix-free GMG solver following the hierarchical structure and smoothing strategies of \cite{dick2011real}.

\paragraph{Matrix-Free EBE Formulation}
To circumvent the overhead of explicit sparse matrix storage and the irregular memory access patterns of global assembly, all linear operators are implemented in a matrix-free \emph{Element-by-Element (EBE)} form~\cite{hughes1983element}. Instead of constructing a global stiffness matrix $\mathbf{K}$, operator applications are computed on-the-fly via tensorized operations:
\begin{equation}
    \mathbf{K}\mathbf{u} \equiv \sum_e \mathbf{A}_e^\top \mathbf{K}_e \mathbf{A}_e \mathbf{u}_e,
\end{equation}
where $\mathbf{K}_e \in \mathbb{R}^{24\times24}$ is the local stiffness matrix (8 nodes $\times$ 3 DoFs), $\mathbf{u}_e$ maps global DoFs to element DoFs, and $\mathbf{A}_e^\top$ scatters contributions back.
Crucially, this formulation aligns strictly with the massive parallelism of modern GPU architectures. By avoiding sparse format (\eg CSR) indirection, it enables coalesced memory access and efficient vectorization, significantly accelerating residual evaluation and smoothing within the neural framework.

\paragraph{Parallel Smoothing}
To dampen high-frequency errors efficiently on GPUs, we employ an \emph{8-color Gauss-Seidel} schedule~\cite{zhang2023}. Nodes $\mathbf{x}_i = (x_i,y_i,z_i)$ are assigned color IDs by spatial parity, ensuring independence within each color class. For each color $c \in \{0,\dots,7\}$, nodes in $\mathcal{C}_c$ are updated in parallel:
\begin{equation}
    \mathbf{u}^{(l)}_i \leftarrow \mathbf{u}^{(l)}_i + \omega (\mathbf{D}^{(l)}_i)^{-1} \mathbf{r}^{(l)}_i, \quad i \in \mathcal{C}_c,
\end{equation}
where $\mathbf{r}^{(l)}$ is the EBE residual, $\omega$ is the relaxation factor, and $\mathbf{D}^{(l)}$ is the diagonal of $\mathbf{K}^{(l)}$. This multi-color sweep provides local relaxation while maintaining full GPU utilization.

\paragraph{Operator Consistency via Galerkin Projection}
To couple grid levels $\{\Omega_l\}_{l=1}^L$, we use topology-consistent trilinear transfer operators: $\mathbf{P}$ (prolongation) and $\mathbf{R} = \mathbf{P}^\mathsf{T}$ (restriction). Coarse-level operators are constructed via local Galerkin projection to preserve physics:
\begin{equation}
    \mathbf{K}_c = \mathbf{R}_{\text{loc}} \, \mathbf{K}_{\text{patch}} \, \mathbf{P}_{\text{loc}},
\end{equation}
where $\mathbf{K}_{\text{patch}}$ aggregates a $2\times2\times2$ patch of fine elements, and $\mathbf{P}_{\text{loc}}$ is the local prolongation matrix. This ensures solver-aware consistency and strict alignment with the neural feature hierarchy in \name. The formulation supports batched execution and periodic wrapping, as detailed in \arxiv{\cref{appendix:E}}{Suppl.~Sec.~5}.

%% file: figures/pipeline.tex
\begin{figure}[t]
    \centering
    \includegraphics[width=\linewidth]{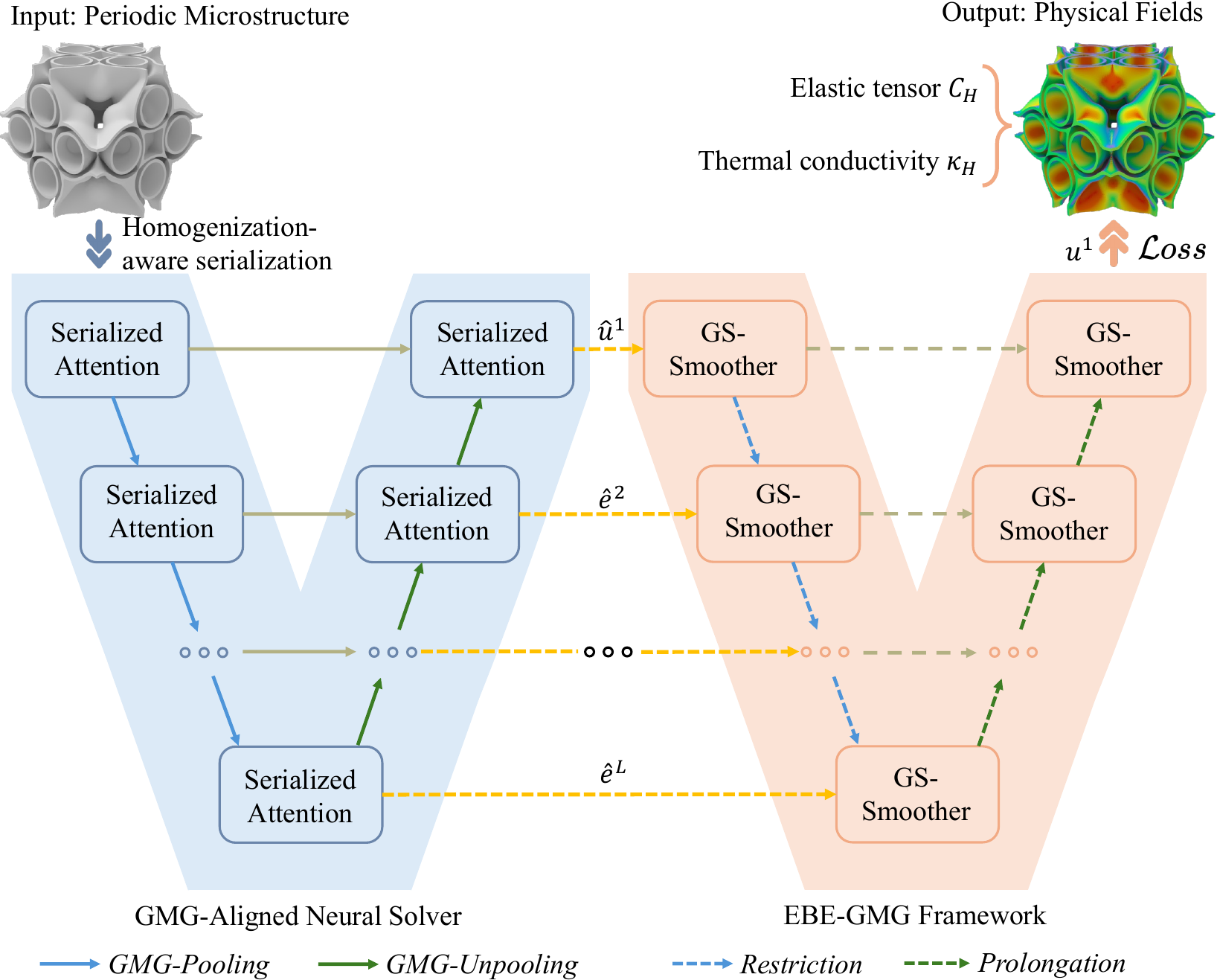}
    \caption{\label{fig:pipeline}
        \textbf{\name framework.}
        \textbf{Left:} The sparse GMG-aligned Transformer backbone featuring homogenization-aware attention and GMG-aware pooling and unpooling.
        \textbf{Right:} The network predicts an initial solution $\hat{u}^1$ and hierarchical error corrections $\hat{e}^l$, which warm-start a differentiable element-by-element GMG V-cycle to produce the refined solution $\mathbf{u}^1$.
    }
\end{figure}

%% file: figures/Isotropic-SA.tex
\begin{figure}[t]
    \centering
    \includegraphics[width=.85\linewidth]{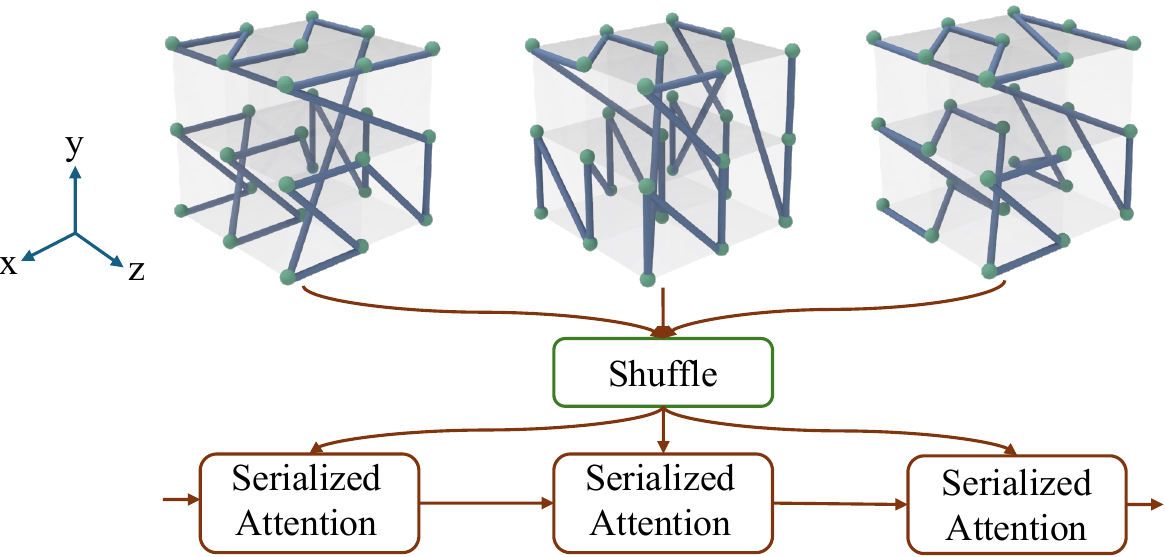}

\caption{\label{fig:IsotropicSA}
\textbf{Homogenization-aware serialized attention.}
The figure illustrates the three complementary Morton curve traversals formed via cyclic coordinate permutations to ensure isotropic neighborhood coverage.
}
\end{figure}

%% file: figures/pooling.tex
\begin{figure}[t]
    \centering
    \includegraphics[width=\linewidth]{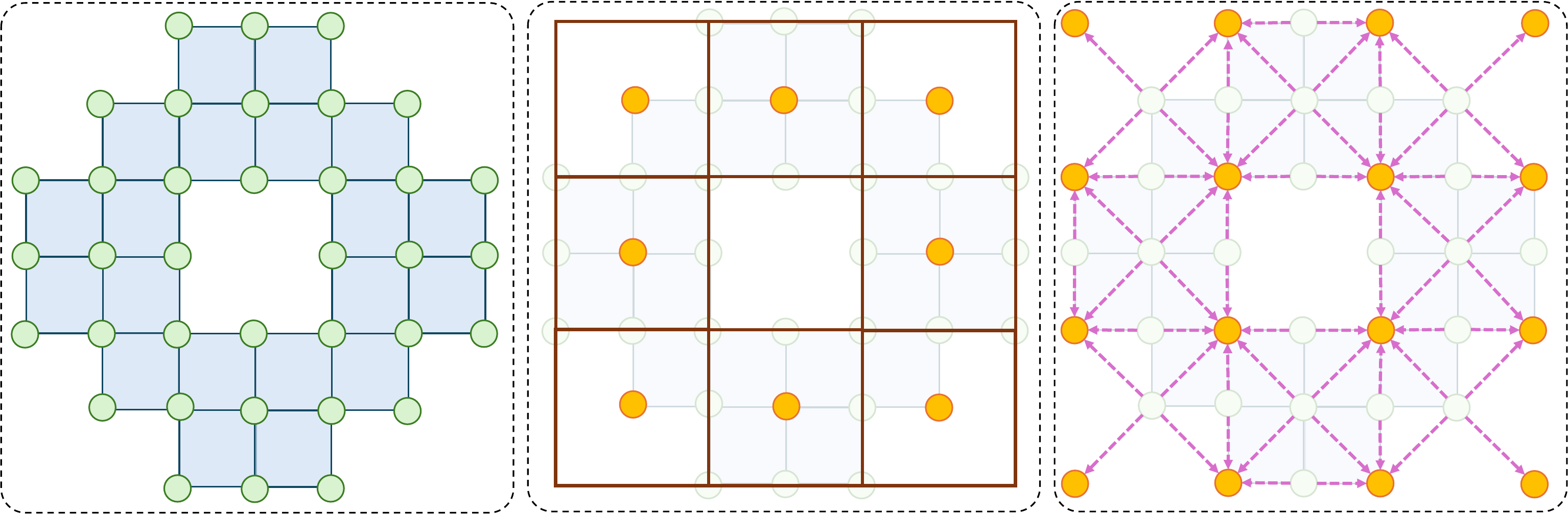}
    \leftline{
        \small
        \hspace{0.07\linewidth}
        (a) Fine grid
        \hspace{0.13\linewidth}
        (b) Voxel Pooling
        \hspace{0.08\linewidth}
        (c) GMG Pooling}
    \caption{
        \textbf{Pooling strategies.}
        (a) Fine-grid layout, where green dots represent nodes at the finer level.
        (b) Standard voxel pooling based on spatial bins and (c) GMG-aware pooling following the multigrid hierarchy.
        In both (b) and (c), the orange dots denote the resulting nodes at the coarser level.
    }
    \label{fig:pooling}

\end{figure}

%% file: src/5-implementation.tex
\section{Experimental Setup}
\label{Implementation}

\subsection{Datasets}

\paragraph{Training Datasets}
The training set contains \num{50000} periodic microstructures~\cite{xing2025pcg}, balanced across three structure classes:
\begin{enumerate}[leftmargin=*]\setlength\itemsep{1mm}
    \item[-] \textbf{Triply Periodic Minimal Surfaces (TPMS)}: \num{10000} samples with smooth, high-curvature interfaces.
    \item[-] \textbf{Parametric Shell Lattices (PSL)}~\cite{liu2022parametric}: \num{12500} samples with shell lattice structures.
    \item[-] \textbf{Truss-like Structures (Truss)}~\cite{zheng2023unifying}: \num{27500} samples with diverse truss-like connectivity.
\end{enumerate}

\paragraph{Periodic Test Sets} We evaluate on three held-out periodic test sets (TPMS, PSL, Truss), each with \num{3000} samples, to benchmark residual convergence and effective property prediction against numerical and neural baselines.
In addition, we include 100 periodic L-BOM microstructures~\cite{wang2026data} as an out-of-distribution (OOD) periodic benchmark with a bicontinuous multiscale topology.

\paragraph{Non-Periodic Test Sets} We benchmark \name's numerical robustness when solving open-boundary linear systems using the Sto-MS dataset~\cite{Martvoronoifoams2016} (\num{1000} non-periodic samples). To accommodate these geometries where standard homogenization is undefined, we adapt the pretrained model by replacing periodic Ra-RoPE with standard RoPE~\cite{su2024roformer} during inference. This effectively removes the toroidal topology without retraining, and is used exclusively for the generalization experiment in~\cref{subsec:accuracy}.

Additional dataset details are provided in \arxiv{\cref{appendix:C}}{Suppl.~Sec.~3}.

\subsection{\name Evaluation Setup}
\label{subsec:metrics}

\paragraph{Evaluation Metrics}
We evaluate \name against  numerical and neural baselines using two metrics:
\begin{enumerate}[leftmargin=*]\setlength\itemsep{1mm}
    \item \textbf{Relative Residual} ($r$):
          $
              r = \frac{\|\mathbf{f} - \mathbf{K}\mathbf{u}\|_2}{\|\mathbf{f}\|_2}
          $,
          where $\mathbf{K}$ is the stiffness or conductivity matrix and $\mathbf{f}$ the load vector. A target $r \leq 10^{-4}$ is used as an engineering-grade stability threshold.
    \item \textbf{Relative Property Error} ($\delta$):
          $
              \delta = \frac{\big|\Phi_{\text{\name}} - \Phi_{\text{GT}}\big|}{\big|\Phi_{\text{GT}}\big|}$,
          where $\Phi_{\text{\name}}$ denotes the effective property computed by \name, and $\Phi_{\text{GT}}$ is the ground truth obtained by running the AMG solver to strict convergence ($r < 10^{-10}$). We report both the mean relative error ($\delta_{\text{mean}}$) and the maximum error ($\delta_{\text{max}}$) across the test set.
\end{enumerate}

\paragraph{Configuration}
All evaluations and comparisons are conducted on a Linux server equipped with 128 CPU cores (\texttt{Intel\textsuperscript{\textregistered} Xeon\textsuperscript{\textregistered} Gold 6530}) and a single NVIDIA RTX 5090 GPU (32\,GB).
Network configurations and training details are provided in \arxiv{\cref{appendix:A}}{Suppl.~Sec.~1}.

%% file: src/6-result.tex
\section{Experimental Results}
\label{sec:results}


\subsection{Efficiency and Scalability}
\input{tables/time-baseline}
We benchmark the end-to-end runtime of \name on periodic microstructure homogenization. Each query requires solving multiple cell problems (six canonical loading modes for linear elasticity).

\paragraph{Comparison to State-of-the-Art Classical Solvers}
\name delivers substantial acceleration compared to established numerical solvers. We benchmark against CPU-based AMGCL~\cite{Demidov2019} (configured with rigid-body null-space modes), and a comprehensive suite of GPU-based baselines: a GPU implementation of preconditioned conjugate gradient (GPU-PCG), NVIDIA AmgX~\cite{naumov2015amgx} (a state-of-the-art algebraic multigrid library), and a specialized Geometric Multigrid (GMG) solver~\cite{zhang2023} optimized explicitly for lattice homogenization.
In~\cref{tab:time}, even at a base resolution of $64^3$, \name is approximately 33$\times$ faster than the fastest GPU baseline. This advantage becomes significantly more pronounced at high resolutions: \name retains its advantage at high resolutions and achieves over $160\times$ speedup relative to the optimized GPU-GMG solver at $512^3$.

This scalability validates our spectral complementarity: while traditional solvers struggle to propagate global low-frequency corrections, \name bypasses this process by predicting a spectrally-aligned initialization. This reduces the problem to a local high-frequency cleanup, resolvable by a single V-cycle across all evaluated resolutions.
Note that to eliminate throughput advantages from parallel batching, \name is evaluated here with a batch size of 1. For a detailed runtime decomposition of both the baseline solvers and the internal components of \name, please refer to \arxiv{\cref{appendix:B}}{Suppl.~Sec.~2} and \arxiv{\cref{appendix:D3}}{Suppl.~Sec.~4.3}.

\paragraph{Analysis of Convergence Behavior}

To better understand the source of the speedup, \cref{fig:res-time} visualizes the wall-clock convergence history of the optimized GPU-GMG solver compared to \name at high resolutions ($256^3$ and $512^3$).
The curves clearly illustrate the fundamental difference in computational paradigms. The classical GMG solver follows a standard iterative trajectory, requiring a substantial accumulation of V-cycles to slowly reduce the residual. This process manifests as a long, descending curve, translating to significant wall-clock time at $512^3$. In contrast, \name achieves this target accuracy almost instantaneously.
In the plot, \name reaches a residual comparable to the converged iterative solver but much earlier along the time axis. This indicates that \name produces an initialization already in the near-converged regime, avoiding the early-stage refinement required by a zero-initialized GMG solve.

\input{figures/time-r}
\input{tables/hybrid-baseline}

\paragraph{Comparison with Neural-Numerical Hybrids}
\name also demonstrates a significant performance margin over recent hybrid PDE solvers, as detailed in~\cref{tab:Hybrid-Baselines}. Existing hybrids typically face a practical trilemma: they fail to simultaneously achieve generalization, convergence, and speed. For instance, methods like GNP~\cite{chen2024graph} and Solver-in-the-Loop (SitL~\cite{um2020solver}) lack generalization, necessitating costly per-structure retraining that renders them unsuitable for real-time design. Operator-replacement methods such as GNN-AMG~\cite{luz2020learning} struggle with stability on high-contrast lattices, often failing to reach solver-grade residuals ($r \approx 10^{-5}$). While CGiNS~\cite{xing2025pcg} offers fast inference, it relies on a truncated PCG schedule that lacks hierarchical error correction, leaving significant residual error ( $\approx 10^{-3}$). In contrast, \name stands out as the first method (to our knowledge) to achieve efficient inference (\SI{0.048}{\second}) and solver-grade accuracy without per-geometry retraining.
These results suggest that our architectural alignment is more effective than loose coupling. By mirroring the numerical solver's topology, \name ensures full-spectrum error correction, enabling a real-time, high-fidelity homogenization engine.

\subsection{Accuracy and Generalization}
We extend our evaluation from computational efficiency to a rigorous assessment of physical fidelity. A viable engineering solver must not only predict effective properties but also ensure the underlying fields satisfy the governing partial differential equations (PDEs) to a high degree of precision—specifically, achieving "solver-grade" residuals ($\approx 10^{-5}$).
\label{subsec:accuracy}
\input{figures/error-distribution}
\input{tables/network-baseline}
\paragraph{Cross-Physics Consistency}
First, we demonstrate the universality of the \name architecture. We evaluate \name on two distinct elliptic PDE systems: vector-valued Linear Elasticity and scalar Steady Heat Conduction over the three periodic test sets (TPMS/PSL/Truss). As shown in~\cref{fig:error-distribution}, \name is not overfitted to a specific operator; instead, it successfully internalizes the generalized multigrid correction mechanism. Across both domains, \name consistently drives relative residuals to the $10^{-5}$ scale while maintaining property errors on the order of $0.01\%$. This cross-physics consistency confirms that our "GMG-aligned" strategy serves as a robust neural corrector independent of the specific physical variable dimension. Furthermore, we demonstrate in \arxiv{\cref{appendix:D2}}{Suppl.~Sec.~4.2} that \name is not limited to the $10^{-5}$ regime; it can be leveraged to achieve even higher numerical precision through iterative refinement or alternative multigrid schedules.

\paragraph{Mechanical Homogenization Benchmark: The Precision Gap}
\cref{tab:network-baseline} benchmarks \name against representative neural homogenization baselines. To diagnose the sources of error, we categorize these baselines into two paradigms:
\begin{enumerate}[leftmargin=*]\setlength\itemsep{1mm}
    \item \emph{Direct Regressors} (\eg 3D-CNN~\cite{rao2020three} ): These methods map geometry directly to properties. Lacking physical constraints (inductive bias), they fail to enforce equilibrium, resulting in unacceptably high errors on complex topologies like PSL and Truss.
    \item \emph{Field Predictors} (\eg Label-Free~\cite{zhu2024learning}, PH-Net~\cite{peng2022ph}, and CGiNS~\cite{xing2025pcg}): These methods predict displacement fields before computing properties. While physically grounded, they suffer from spectral bias: without rigorous numerical correction, the predicted fields ``drift'' from the true equilibrium, leaving high residual errors ($10^{-3}$ to $10^{-2}$).
\end{enumerate}

\paragraph{Architectural vs. Spectral Alignment}
Even the strongest baseline, CGiNS~\cite{xing2025pcg}, which employs a differentiable PCG solver (``numerical-spectral alignment''), remains constrained by the limited convergence depth of single-level Krylov methods. In contrast, \name achieves architectural alignment: by embedding the neural predictor within a GMG V-cycle, we explicitly target and eliminate error modes across the entire frequency spectrum.
Quantitatively, this yields a substantial improvement in performance. On the challenging Truss dataset, \name reduces the mean relative property error from $0.51\%$ (CGiNS) to $0.05\%$ -- an order-of-magnitude improvement -- while simultaneously driving the mean residual from $5.50 \times 10^{-3}$ down to $2.44 \times 10^{-5}$.

\input{figures/multigrid}
\paragraph{Robustness on Out-of-Distribution Geometries}
Finally, we stress-test \name's generalization capabilities on geometries unseen during training:
\begin{enumerate}[leftmargin=*]\setlength\itemsep{1mm}
    \item \emph{Unseen Periodic Topologies}: On the L-BOM~\cite{wang2026data} dataset, \name maintains its solver-grade performance ($\delta_{Mean} = 0.02\%$, $r_{Mean} \approx 4.5 \times 10^{-5}$), proving the model has learned the homogenization operator rather than memorizing geometric features.
    \item \emph{Non-Periodic/Stochastic Structures}: We use the Sto-MS dataset to evaluate robustness on open-boundary topologies. Aside from the absence of geometric periodicity, the experimental configuration is kept identical to the standard homogenization setting. Although these systems arise from non-periodic structures outside the training distribution, the pretrained \name model remains effective, reducing the FEM linear system residual to $7.44 \times 10^{-5}$ without periodic constraints. This demonstrates that the "Neural + GMG Correction" paradigm maintains high numerical stability on irregular geometries, effectively generalizing beyond the periodic settings for which it was trained.
\end{enumerate}

\subsection{Modular Coupling with Multigrid: Schedules and Smoothers}

\paragraph{Schedule Analysis: The Accuracy-Efficiency Trade-off}
\name is designed not merely as a rigid black-box solver, but as a composable operator. By decoupling the learned Serialized Attention modules (blue nodes) from the numerical smoothers (yellow nodes), \name can be seamlessly rewired to instantiate a broad family of multigrid schedules as shown in~\cref{fig:multigrid}.

Leveraging this modularity, we probe the fundamental interplay between neural prediction and numerical relaxation on the $128^3$-resolution  Truss benchmark (\cref{tab:GMG-generalization}). The results reveal a distinct trade-off governed by the recursive structure and smoothing budget:
\begin{enumerate}[leftmargin=*]\setlength\itemsep{1mm}
    \item \emph{The W-Cycle}: By performing more aggressive coarse-grid corrections, the W-cycle achieves the lowest terminal residual (on the order of $10^{-6}$). However, this comes at the highest computational cost (\SI{0.152}{\second}).
    \item \emph{The V-Cycle}: The standard V-cycle represents the optimal ``sweet spot'' delivering solver-grade accuracy (on the order of $10^{-5}$) with significantly lower latency (\SI{0.123}{\second}).
    \item \emph{The Necessity of Relaxation}: Notably, schedules with truncated relaxation stages, such as Half-V or Full Multigrid (FMG) variants, exhibit degraded residuals ($10^{-4}$ range).
\end{enumerate}
This empirically validates our hybrid philosophy: even with a potent neural initialization, sufficient numerical relaxation remains indispensable for damping the high-frequency error components that neural networks struggle to resolve.
\input{tables/GMG-generalization}

\paragraph{Smoother Analysis: Spectral Alignment vs. Over-Structuring}
We further investigate the impact of the numerical smoother choice in~\cref{tab:numerical-solvers}. This ablation yields an instructive finding: more complex solvers do not imply better smoothers. While one might expect powerful Krylov subspace methods (like CG or PCG) to outperform simple stationary iterations, our results show the opposite. Classical 8-color Gauss-Seidel (G-S) and SOR achieve the best convergence ($2.44 \times 10^{-5}$), whereas replacing the smoother with CG significantly degrades performance ($2.76 \times 10^{-4}$).

This phenomenon is a classic case of spectral misalignment. The role of a smoother in GMG is not to minimize global error, but to efficiently damp local, high-frequency oscillations to prepare the residual for coarse-grid correction. Krylov methods, which optimize over a global subspace, are ``computationally over-structured'' for this specific local task. Under the limited iteration budget of a smoothing step, they fail to target high-frequency modes as effectively as G-S, leading to inferior overall convergence per unit of compute. This confirms that \name's success relies on respecting the spectral roles of its components: the neural network handles global low-frequencies, while classical smoothers handle local high-frequencies.

\subsection{Ablation Studies: Deconstructing Architectural Alignment}
\label{subsec:ablation}
\input{tables/solver-generalization}
\input{tables/combine-ablation}

To rigorously validate the necessity of each \name component, we conduct a two-stage ablation study: first dissecting the macro-level coupling strategy and then isolating the micro-level modules.

\paragraph{Macro-Ablation: Backbone and Solver Integration}
\cref{tab:ablation-architecture} investigates the interplay between the neural backbone (Sparse CNN vs. Transformer) and the numerical coupling mechanism.
\begin{enumerate}[leftmargin=*]\setlength\itemsep{1mm}
    \item \emph{The Power of Long-Range Attention}: Under identical coupling conditions, the Transformer backbone consistently outperforms the CNN (\eg $7.70 \times 10^{-4}$ vs. $8.12 \times 10^{-4}$ with single-level GMG). This confirms that attention mechanisms are superior at capturing the global, long-range elastic couplings that dominate the low-frequency error spectrum—precisely the errors that local convolutions struggle to resolve.
    \item \emph{The Insufficiency of Loose Spectral Alignment}: We first evaluate a loose coupling strategy (``Network + PCG''), where the neural network provides a spectrally-aligned initialization to a generic Krylov solver. Despite the improved starting point, this approach stagnates at residuals of $10^{-3}$ level. This failure proves that merely appending a linear solver is insufficient; without a geometric hierarchy to efficiently resolve the remaining error spectrum, the solver cannot capitalize on the neural prediction.
    \item \emph{The Necessity of Structural Alignment}: The decisive performance leap stems from structural alignment—matching the neural architecture to the solver's topology. While injecting predictions only at the finest level (``Weak Coupling'') yields residuals around $10^{-4}$, the ``Strong Coupling'' (GMG-multi) approach—where the network explicitly predicts corrections for every level of the multigrid hierarchy—reduces the residual by another order of magnitude to $2.44 \times 10^{-5}$. This shows that the neural network must be structurally pervasive, guiding the numerical relaxation at every scale rather than merely providing a flat initialization.
\end{enumerate}

\input{tables/ablation-components}
\paragraph{Micro-Ablation: Physics-Aware Components}
\cref{tab:ablation_components} isolates the impact of \name’s three custom modules designed to enforce physical consistency.
\begin{enumerate}[leftmargin=*]\setlength\itemsep{1mm}
    \item \emph{GMG-Aware vs. Voxel Pooling}: Replacing our GMG-aware pooling with standard spatial-binning (Voxel Pooling) degrades the residual by nearly 40$\times$ ($1.01 \times 10^{-3}$). Standard pooling ignores the strict parent-child node relationships required by the numerical grid, breaking the mathematical validity of the coarse-grid correction. Our method ensures the neural hierarchy is topologically isomorphic to the solver's restriction operator.
    \item \emph{Homogenization-Aware Attention}: Reverting to standard PTv3 serialization increases residuals to $7.21 \times 10^{-4}$. Standard space-filling curves introduce ``index jumps'' that create anisotropic blind spots in the receptive field. Our multi-view cyclic serialization ensures isotropic neighborhood coverage, which is critical for capturing the symmetric nature of physical stress propagation in periodic lattices.
    \item \emph{Ra-RoPE vs. RoPE}: Replacing our Ra-RoPE with standard RoPE degrades performance to $7.41 \times 10^{-4}$. Standard RoPE treats periodic boundaries ($x=0$ and $x=\Upsilon$) as distant. In contrast, Ra-RoPE parameterizes rotation by the physical period, enforcing phase continuity across boundaries. This ensures that the network respects the PBCs of the periodic lattice, a non-negotiable constraint for accurate homogenization.
\end{enumerate}

%% file: tables/time-baseline.tex
\begin{table}[t]
    \centering
    \caption{\label{tab:iso-accuracy-runtime}
        \textbf{Iso-accuracy runtime comparison.}
        Average wall-clock time (seconds) for end-to-end homogenization under a strict iso-accuracy stopping criterion.
        For each query, all numerical baselines are run until they reach the same relative residual achieved by \name.
        \name is evaluated with a batch size of one.
    }
    \scalebox{0.85}{
        \begin{tabular}{cc|rrrrr }
            \toprule
            $\bf{N_{res}}$                      &
            $ \bm{r} $ ($\times 10^{-5}$)       &
            \multicolumn{1}{c}{\texttt{AMGCL}}  &
            \multicolumn{1}{c}{\texttt{AmgX } } &
            \multicolumn{1}{c}{\texttt{GMG } }  &
            \multicolumn{1}{c}{\texttt{PCG } }  &
            \multicolumn{1}{c}{\textbf{\name }}                                                                                    \\
            \midrule
            64                                  & 5.62 & 94.78   & \underline{1.63} & 3.33               & 2.91   & \textbf{0.048} \\
            128                                 & 7.23 & 118.52  & 14.95            & \underline{12.11}  & 22.23  & \textbf{0.123} \\
            256                                 & 4.67 & 482.73  & 68.92            & \underline{50.64}  & 88.43  & \textbf{0.595} \\
            512                                 & 7.58 & 1138.52 & 441.52           & \underline{389.29} & 623.27 & \textbf{2.378} \\
            \bottomrule
        \end{tabular}
    }
    \label{tab:time}
\end{table}

%% file: figures/time-r.tex
\begin{figure}[t]
    \centering
    \includegraphics[width=\linewidth]{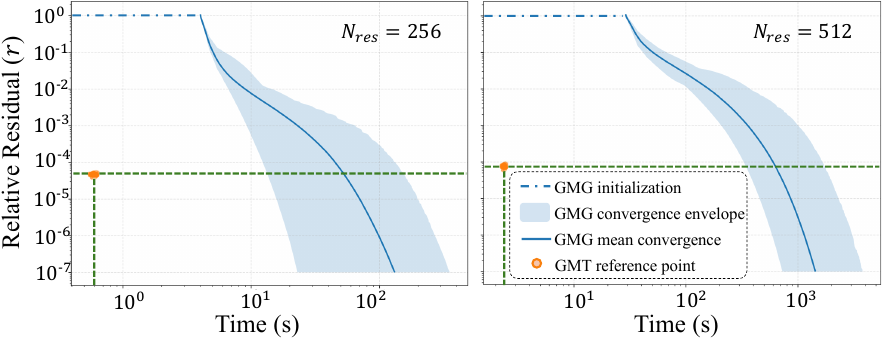}
    \caption{\label{fig:res-time}
        \textbf{Wall-clock convergence comparison at high resolutions.}
        Mean relative residual versus wall-clock time (seconds) for solving all six canonical strain cases.
        \textbf{Left:} $256^3$.
        \textbf{Right:} $512^3$.
        Solid lines show the iterative V-cycle convergence of the classical GMG solver, while the marker indicates a single inference pass of \name.
    }
\end{figure} 

%% file: tables/hybrid-baseline.tex
\begin{table}[t]
    \caption{\label{tab:Hybrid-Baselines} \textbf{Comparison with state-of-the-art neural-numerical hybrid solvers.} Methods are categorized by coupling strategy: Learned Preconditioner (GNP~\cite{chen2024graph}), Operator Replacement (GNN-AMG~\cite{luz2020learning}), and Learned Initial Guess (SitL~\cite{um2020solver}, CGiNS~\cite{xing2025pcg}). Generalization: Transferability to unseen microstructures without retraining. Convergence: Guaranteed monotonic error reduction during subsequent iterative refinement, avoiding stagnation or divergence. Pre-Time / For-Time: Distinguishes per-structure precomputation costs from test-time inference latency. 
    }
    \scalebox{0.7}{
        \begin{tabular}{cccccc}
            \toprule

            \textbf{Network}        & \textbf{GNP}                             & \textbf{SitL}                            & \textbf{GNN-AMG}                         & \textbf{CGiNS}                           & \textbf{\name}                                \\
            \midrule
            \textbf{Type}           & Precond.                                 & Initial Guess                 & Operator                                 & Initial Guess                 & Initial Guess                      \\
            \textbf{Generalization} & \xmark                                   & \xmark                                   & \cmark                                   & \cmark                                   & \cmark                                        \\
            \textbf{Convergence}    & \cmark                                   & \cmark                                   & \xmark                                   & \cmark                                   & \cmark                                        \\
            \textbf{Pre-Time (s)}   & \multicolumn{1}{c}{643}                  & \multicolumn{1}{c}{14960}                & --                                       & --                                       & --                                            \\
            \textbf{For-Time (s)}   & 4.23                                     & \multicolumn{1}{c}{1.03}                 & 1.36                                     & 0.15                                     & \multicolumn{1}{c}{\textbf{0.05}}             \\
            $\bm{ r_{mean}}$        & \multicolumn{1}{r}{$3.89\times10^{-4}$ } & \multicolumn{1}{r}{$1.33\times10^{-5}$ } & \multicolumn{1}{r}{$4.92\times10^{-3}$ } & \multicolumn{1}{r}{$3.46\times10^{-3}$ } & \multicolumn{1}{r}{$\bf{2.44\times10^{-5}}$ } \\
            \bottomrule
        \end{tabular}
    }
\end{table}

%% file: figures/error-distribution.tex
\begin{figure}[t]
    \centering
    \includegraphics[width=\linewidth]{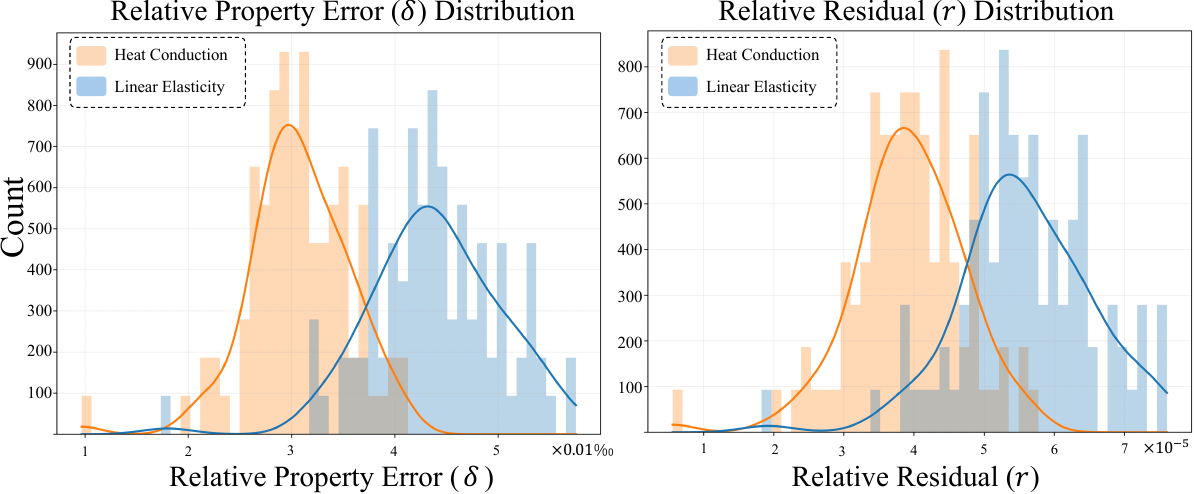}
    \caption{\label{fig:error-distribution}
        Distributions of relative property error $\delta$ (left) and relative residual $r$ (right) for linear elasticity and steady heat conduction evaluated on periodic lattices (TPMS, PSL, Truss).
    }
\end{figure}

%% file: tables/network-baseline.tex
\begin{table}[t]
    \caption{\label{tab:network-baseline}
        \textbf{Mechanical homogenization accuracy benchmark.}
        Comparison of \name with direct regression (3D-CNN~\cite{rao2020three}) and field-prediction baselines (Label-Free~\cite{zhu2024learning}, PH-Net~\cite{peng2022ph}, CGiNS~\cite{xing2025pcg}) across five test sets.
        We report the mean relative property error ($\delta$) and the mean relative residual ($r$). 3D-CNN directly regresses effective properties and reports only $\delta$.
        For the non-periodic Sto-MS dataset, homogenization is inapplicable and only $r$ is reported.
    }

    \scalebox{0.85}{
        \begin{tabular}{llrrrr}
            \toprule
            \textbf{Testset}       & \multicolumn{1}{c}{\textbf{ Network}} & \multicolumn{1}{c}{$\bm{\delta_{max}}$} & \multicolumn{1}{c}{$\bm{\delta_{mean}}$} & \multicolumn{1}{c}{$\bm{ r_{max}}$} & \multicolumn{1}{c}{$\bm{ r_{mean}}$} \\
            \midrule
            \multirow{5}{*}{TPMS}  &
            3D-CNN                 & $17.57\%$                             & $8.54\%$                                & \multicolumn{1}{c}{n.a.}                 & \multicolumn{1}{c}{n.a.}                                                   \\
                                   & Label-Free                            & $44.81\%$                               & $23.51\%$                                & $6.83\times10^{-2}$                 & $8.76\times10^{-2}$                  \\
                                   & PH-Net                                & $2.62\%$                                & $1.45\%$                                 & $6.73\times10^{-2}$                 & $2.14\times10^{-2}$                  \\
                                   & CGiNS                                 & $0.18\%$                                & $0.05\%$                                 & $3.02\times10^{-3}$                 & $2.54\times10^{-3}$                  \\
                                   & \textbf{\name}                        & $\bf{0.05}$\textperthousand             & $\bf{0.03}$\textperthousand              & $\bf{8.94\times10^{-5}}$            & $\bf{6.95\times10^{-5}}$             \\
            \midrule
            \multirow{5}{*}{PSL}   &
            3D-CNN                 & $111.99\%$                            & $40.16\%$                               & \multicolumn{1}{c}{n.a.}                 & \multicolumn{1}{c}{n.a.}                                                   \\
                                   & Label-Free                            & $239.98\%$                              & $87.51\%$                                & $7.05\times10^{-2}$                 & $6.13\times10^{-2}$                  \\
                                   & PH-Net                                & $47.34\%$                               & $ 23.19\%$                               & $5.04\times10^{-2}$                 & $3.28\times10^{-2}$                  \\
                                   & CGiNS                                 & 0.64\%                                  & 0.31\%                                   & $6.12\times10^{-3}$                 & $4.01\times10^{-3}$                  \\
                                   & \textbf{\name}                        & $\bf{0.04}$\textperthousand             & $\bf{0.03}$\textperthousand              & $\bf{1.00\times10^{-4}}$            & $\bf{7.48\times10^{-5}}$             \\

            \midrule
            \multirow{5}{*}{Truss} &
            3D-CNN                 & $197.54\%$                            & $47.13\%$                               & \multicolumn{1}{c}{n.a.}                 & \multicolumn{1}{c}{n.a.}                                                   \\
                                   & Label-Free                            & $836.25\%$                              & $177.47\%$                               & $6.90\times10^{-2}$                 & $5.65\times10^{-2}$                  \\
                                   & PH-Net                                & $357.14\%$                              & $65.67\%$                                & $3.70\times10^{-2}$                 & $2.45\times10^{-2}$                  \\
                                   & CGiNS                                 & $0.84\%$                                & $0.51\% $                                & $1.02\times10^{-2}$                 & $5.50\times10^{-3}$                  \\
                                   & \textbf{\name}                        & $\bf{0.13}$\textperthousand             & $\bf{0.05}$\textperthousand              & $\bf{2.88\times10^{-5}}$            & $\bf{2.44\times10^{-5}}$             \\
            \midrule
            \multirow{5}{*}{L-BOM} &
            3D-CNN                 & $16.51\%$                             & $7.13\%$                                & \multicolumn{1}{c}{n.a.}                 & \multicolumn{1}{c}{n.a.}                                                   \\
                                   & Label-Free                            & $37.51\%$                               & $22.08\%$                                & $2.71\times10^{-2}$                 & $1.52\times10^{-2}$                  \\
                                   & PH-Net                                & $12.01\%$                               & $2.37\%$                                 & $3.23\times10^{-2}$                 & $1.03\times10^{-2}$                  \\
                                   & CGiNS                                 & $0.27\%$                                & $0.08\%$                                 & $5.40\times10^{-3}$                 & $3.77\times10^{-3}$                  \\
                                   & \textbf{\name}                        & $\bf{0.04}$\textperthousand             & $\bf{0.02}$\textperthousand              & $\bf{8.01\times10^{-5}}$            & $\bf{4.52\times10^{-5}}$             \\
            \midrule
            \multirow{4}{*}{Sto-MS}
                                   & Label-Free                            & \multicolumn{1}{c}{n.a.}                & \multicolumn{1}{c}{n.a.}                 & $8.38\times10^{-1}$                 & $3.53\times10^{-1}$                  \\
                                   & PH-Net                                & \multicolumn{1}{c}{n.a.}                & \multicolumn{1}{c}{n.a.}                 & $3.73\times10^{-2}$                 & $1.61\times10^{-2}$                  \\
                                   & CGiNS                                 & \multicolumn{1}{c}{n.a.}                & \multicolumn{1}{c}{n.a.}                 & $7.66\times 10^{-3}$                & $4.19\times 10^{-3}$                 \\
                                   & \textbf{\name}                        & \multicolumn{1}{c}{n.a.}                & \multicolumn{1}{c}{n.a.}                 & $\bf{9.35\times10^{-5}}$            & $\bf{7.44\times10^{-5}}$             \\
            \bottomrule
        \end{tabular}
    }
\end{table}

%% file: figures/multigrid.tex
\begin{figure}[t]
  \centering
  \includegraphics[width=\linewidth]{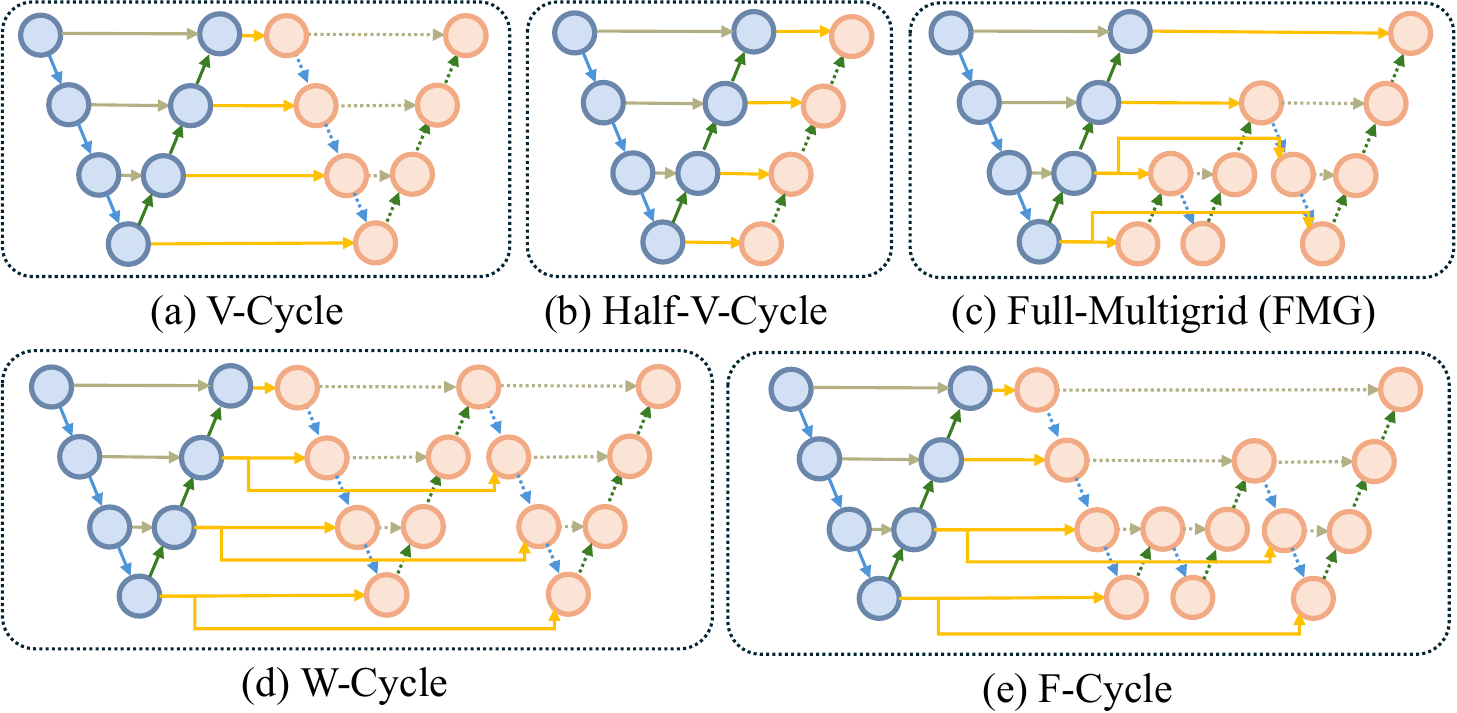}
  \caption{\label{fig:multigrid} \textbf{\name as a plug-in module for different multigrid schedules.}
    Learnable attention modules (blue) and numerical smoothers (yellow) can be flexibly rewired. By rewiring these components, \name can be instantiated as (a) V-cycle, (b) Half-V-cycle, (c) Full Multigrid (FMG), (d) W-cycle, and (e) F-cycle. For schedules with multiple numerical stages, a multi-head decoder is utilized for stage-wise injection.}
\end{figure}

%% file: tables/GMG-generalization.tex
\begin{table}[t]
    \centering
    \caption{\label{tab:GMG-generalization} \textbf{Effect of GMG schedule on \name inference for the 128-resolution Truss benchmark.} We report wall-clock inference time and mean relative residual $r_{mean}$ for coupling \name with different multigrid schedules (V/F/W, Half-V, and Full Multigrid (FMG)).}
    \scalebox{0.85}{
        \begin{tabular}{l|cr}
            \toprule
            \textbf{GMG Type} & \multicolumn{1}{c}{\textbf{Time} (s)} & \multicolumn{1}{c}{$\bm{ r_{mean}}$ } \\
            \midrule
            V-Cycle           & 0.123                                 & $2.44\times10^{-5}$                   \\
            Half-V-Cycle      & 0.105                                 & $2.22\times10^{-4}$                   \\
            FMG               & 0.118                                 & $5.65\times10^{-4}$                   \\
            W-Cycle           & 0.152                                 & $8.13\times10^{-6}$                   \\
            F-Cycle           & 0.130                                 & $3.94\times10^{-5}$                   \\
            \bottomrule
        \end{tabular}
    }
\end{table}

%% file: tables/solver-generalization.tex
\begin{table}[t]
    \centering
    \caption{\label{tab:numerical-solvers}
        \textbf{Impact of numerical smoother choice.}
        Performance comparison of local relaxation methods and Krylov subspace solvers when used as smoothers under a fixed smoothing budget.
    }
    \scalebox{0.85}{
        \begin{tabular}{l|r}
            \toprule
            \textbf{Smoother}    & \multicolumn{1}{c}{$\bm{ r_{mean}}$ } \\
            \midrule
            Jacobi               & $9.04\times10^{-4}$                   \\
            G-S (8-color)        & $2.44\times10^{-5}$                   \\
            SOR ($\omega = 0.8$) & $2.81\times10^{-5}$                   \\
            SOR ($\omega = 1.2$) & $4.77\times10^{-5}$                   \\
            CG                   & $2.76\times10^{-4}$                   \\
            PCG (Jacobi)         & $1.57\times10^{-4}$                   \\
            \bottomrule
        \end{tabular}
    }
\end{table}

%% file: tables/combine-ablation.tex
\begin{table}[t]
    \caption{\label{tab:ablation-architecture}
        \textbf{Backbone and solver-coupling ablation.}
        Comparison of Sparse CNN and Transformer backbones under different solver coupling strategies.
        ``PCG'' denotes loose coupling with a conjugate gradient refiner.
        ``GMG (single)'' injects neural predictions only at the finest level, while
        ``GMG (multi)'' injects hierarchical corrections at all multigrid levels.
    }
    \scalebox{0.85}{
        \begin{tabular}{cl|l}
            \toprule
            \textbf{Network}             & \textbf{Numerical solver} & $\bm{r_{mean}}$          \\
            \midrule
            \multirow{3}{*}{CNN}         & PCG                       & $4.33\times10^{-3}$      \\
                                         & GMG (single)              & $8.12\times10^{-4}$      \\
                                         & GMG (multi)               & $7.70\times10^{-4}$      \\
            \midrule
            \multirow{3}{*}{Transformer} & PCG                       & $2.69\times10^{-3}$      \\
                                         & GMG (single)              & $7.14\times10^{-4}$      \\
                                         & GMG (multi)               & $\bm{2.44\times10^{-5}}$ \\
            \bottomrule
        \end{tabular}
    }
\end{table}

%% file: tables/ablation-components.tex
\begin{table}[t]
    \caption{\label{tab:ablation_components} \textbf{Component-level ablation study.} We isolate the impact of three key physical-aware modules by replacing them with standard deep learning alternatives. \textbf{Standard \name} (bottom row) integrates GMG-Pooling, Homogenization-aware Attention, and Ra-RoPE, achieving orders-of-magnitude lower residuals compared to standard configurations.}
    \centering
    \scalebox{0.85}{
        \begin{tabular}{llc}
            \toprule
            \textbf{Ablated Module}    & \textbf{Configuration Variant} & $\bm{r_{mean}}$                \\ \midrule
            Pooling Strategy           & w/ Voxel Pooling               & $1.01 \times 10^{-3}$          \\
            Attention Mechanism        & w/ PTv3 Attention              & $7.21 \times 10^{-4}$          \\
            Positional Encoding        & w/ RoPE \cite{su2024roformer}  & $7.41 \times 10^{-4}$          \\ \midrule
            \textbf{Full Architecture} & \textbf{Standard \name}        & $\mathbf{2.44 \times 10^{-5}}$ \\
            \bottomrule
        \end{tabular}
    }
\end{table}

%% file: src/7-application.tex
\section{Applications}

\subsection{Real-time Screening for Generative Models}
\input{figures/Generative-model}
Generative models like Diffusion Models are powerful for inverse microstructure design but suffer from stochasticity, necessitating a ``Generate-and-Filter'' pipeline to ensure performance compliance. Traditional numerical homogenization is prohibitively slow for such high-throughput screening.

To bridge this gap, we leverage \name for real-time screening within the MIND (Microstructure INverse Design) framework~\cite{xue2025mind}. As shown in~\cref{fig:Generative}, we demonstrate this on a macro-scale bracket comprising 320 spatially distributed unit cells . For each cell, MIND generates 64 candidates at $N_{res}=64$, totaling 20,480 structures that require physics-based assessment. Using the GPU-accelerated AmgX solver (~1.64s per sample), screening this dataset takes approximately 11 hours. In contrast, \name (batch size 32) completes the entire task in just 4.8 minutes. This speedup of over two orders of magnitude transforms validation from an offline bottleneck into a rapid, interactive design workflow.

\subsection{Accelerating Large-Scale Multi-Scale Simulation}
\input{figures/two-scale}
Many practical lattice-enabled designs go beyond single periodic RVEs, typically manifesting as complex arrayed structures on free-form geometries. We validate \name's capability on a multi-scale topology optimization model, as shown in~\cref{fig:Two-scale}. This macro-structure consists of \num{3204} distinct periodic unit cells. With each lattice structure discretized at $N_{res}=64$, the total number of fine-scale nodes exceeds $2.2 \times 10^8$. Although direct numerical simulation of such a massive system is computationally achievable, the enormous time overhead renders it infeasible for rapid iterative design.

To address this, we employ a decoupled two-scale simulation paradigm. The macro-model is first decomposed into its constituent lattice structures. \name then serves as a high-throughput engine to compute the homogenized effective properties for each of the \num{3204} unique lattices in parallel. These effective properties are subsequently mapped back to the coarse mesh to assemble a macro-scale effective model, which is solved under prescribed boundary conditions.

The efficiency gains are substantial. For the reported model, traditional numerical homogenization takes over 1.5 hours to process the full set of unit cells. In stark contrast, \name completes the entire batch in just \textbf{40 seconds}. This considerable speedup effectively eliminates the computational bottleneck in multi-scale analysis, allowing for the rapid evaluation of billion-node scale metamaterials that were previously intractable.

\subsection{Fast Inverse Homogenization Design}
\label{subsec:app-IHD}
Beyond forward homogenization, \name serves as a high-efficiency differentiable accelerator for inverse topology optimization. Unlike conventional compliance minimization, unit-cell optimization aims to tailor specific effective properties (\eg Young’s modulus $E$, shear modulus $G$, or bulk modulus $K_B$).

During optimization, the discrete voxel indicator is relaxed to a continuous density field $\rho \in [\rho_{min}, 1]$ under the standard SIMP formulation. Accordingly, we use density-valued local features and retrain \name for this task, rather than directly applying the binary pretrained model. We further replace generic reverse-mode autodiff with analytically derived sensitivities with respect to $\rho$, reducing the optimization cost at $64^3$ from $140$ ms/iteration reported in~\cite{zhang2023} to $90$ ms/iteration in our implementation. We note that this speedup appears moderate in part because the GMG solver in~\cite{zhang2023} is terminated at a relative residual of $10^{-2}$, corresponding to a looser solve tolerance, whereas \name operates at a higher homogenization accuracy. Therefore, this timing comparison is conservative and does not fully reflect the advantage of \name at matched solution quality. To maintain numerical efficiency, voxels with $\rho < 10^{-5}$ are pruned every 20 iterations (More details are provided in \arxiv{\cref{appendix:F3}}{Suppl.~Sec.~6.3}).

As shown in~\cref{fig:TO-model}, \name can optimize lattice topologies starting from random noise. Under a fixed volume fraction constraint, the optimizer successfully evolves distinct morphologies specialized for maximizing $E$, $G$, or $K_B$. \name also functions as a data augmentation engine. By using existing high-performing structures from our database as initializations, we can further optimize them to push the boundaries of the achievable property space. \cref{fig:TO-expandEG} demonstrates that this fine-tuning process yields new unit cells that significantly expand the performance envelope of the original dataset.

\input{figures/TO-model}
\input{figures/TO-expand}
\input{figures/pareto}

\subsection{High-Throughput Design Space Exploration and Pareto Front Construction}

We leveraged the developed deep learning framework to perform a high-throughput scan of \num{300000} lattice structures—a dataset explicitly augmented by the inverse design pipeline in~\cref{subsec:app-IHD} to capture extreme performance boundaries. This massive evaluation task is computationally prohibitive for traditional FEM. The variables—effective Young's modulus ($E$), thermal conductivity ($\kappa$), and volume fraction ($v_f$)—form the core triad of multifunctional optimization, balancing structural stability, energy dissipation, and lightweight constraints. Remarkably, the entire dataset was processed in just 2 hours for mechanical properties and 55 minutes for thermal conductivity, representing a speedup of over 100 times.

This dataset serves as a library for multifunctional designs, such as thermo-mechanically coupled satellite components and micro-channel heat sinks for electronics. It also acts as a pre-optimized library for real-time generative design. As shown in~\cref{fig:Pareto}, the resulting 3D Pareto front illustrates the performance boundaries, from which six representative points were identified:

\begin{enumerate}[label=(\alph*)]
    \item \textbf{Absolute Maximum Stiffness}: Represents the stiffness limit ($E = 0.693$, $v_f = 0.844$), ideal for heavy-duty components where stiffness is the sole priority.
    \item \textbf{Specific Stiffness Champion}: Features the highest $E / v_f$ ratio, defining maximum load-bearing efficiency per unit mass for weight-sensitive aerospace applications.
    \item \textbf{Low-density Thermal Priority}: Prioritizes heat dissipation ($\kappa = 0.071$) at low density ($v_f < 15\%$), revealing efficient sparse conduction paths.
    \item \textbf{Global Knee Point}: The best compromise between all three objectives, located closest to the "ideal point" in the performance space.
    \item \textbf{Mid-density Interval Knee}: Selected from the $0.15 < v_f < 0.3$ range, this point aligns with the optimal processing windows for industrial additive manufacturing (\eg SLM, SLA).
    \item \textbf{Low-density Comprehensive Champion}: Demonstrates the strongest synergistic $E$-$\kappa$ performance under low-density constraints ($v_f < 0.15 $).
\end{enumerate}

%% file: figures/Generative-model.tex
\begin{figure*}[t]
  \centering
\includegraphics[width=.9\linewidth]{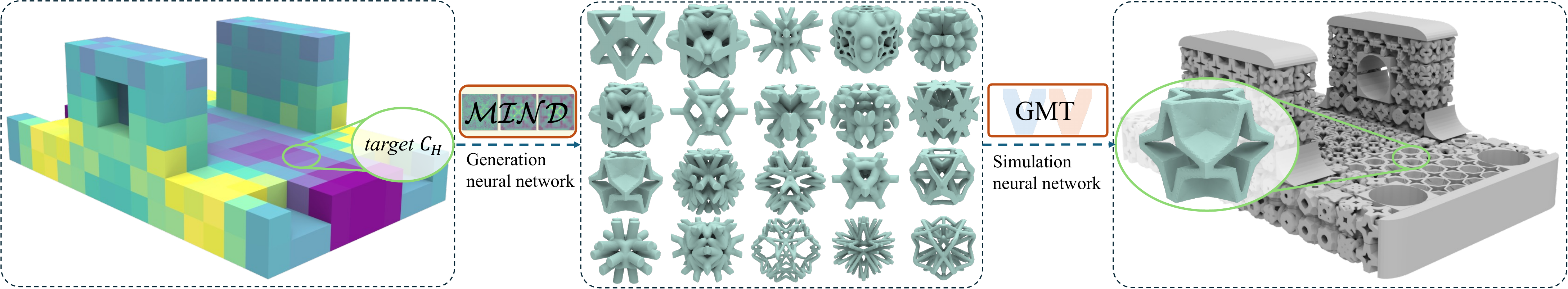}
\leftline{ 
        \hspace{0.13\linewidth}
        (a) 
         \hspace{0.325\linewidth}
        (b) 
        \hspace{0.33\linewidth}
        (c) }

\caption{\label{fig:Generative}
The ``Generate-and-Filter'' workflow for inverse design.
(a) A macro-scale bracket with region-wise stiffness targets.
(b) Multiple lattice candidates generated per target by the MIND model.
(c) High-throughput screening of candidates using \name.
}
\end{figure*}

%% file: figures/two-scale.tex
\begin{figure}[t]
  \centering
  \includegraphics[width=.9\linewidth]{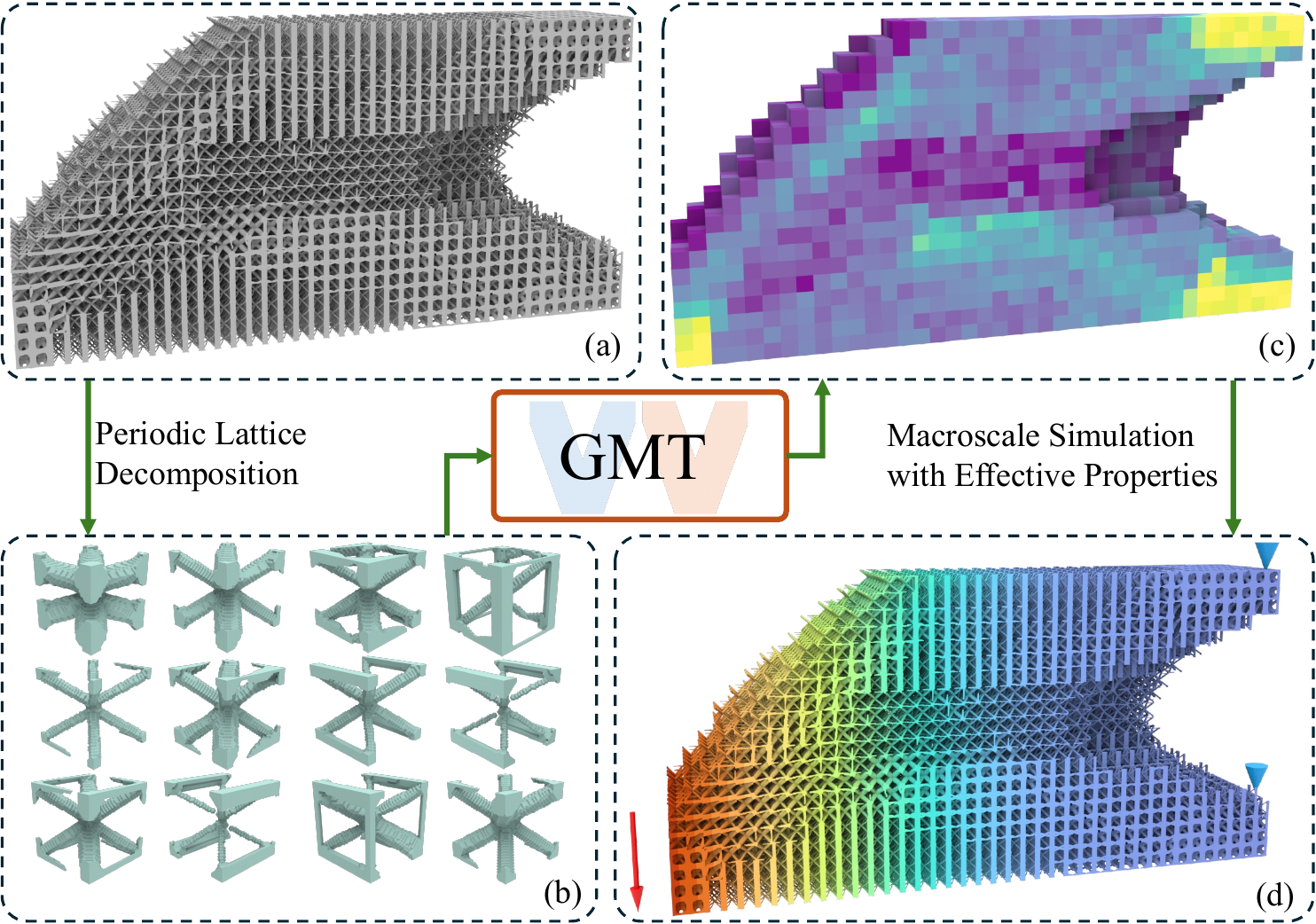}

  \caption{\label{fig:Two-scale}
    \textbf{Two-scale simulation pipeline.}
    (a) A multi-scale structure composed of spatially varying periodic unit cells.
    (b) Decomposition into individual lattice domains.
    (c) Homogenization of all unit cells to obtain a spatially varying effective material map.
    (d) Macro-scale simulation using the homogenized properties to predict structural deformation.
  }
\end{figure}

%% file: figures/TO-model.tex
\begin{figure}[t]
  \centering
  \includegraphics[width=\linewidth]{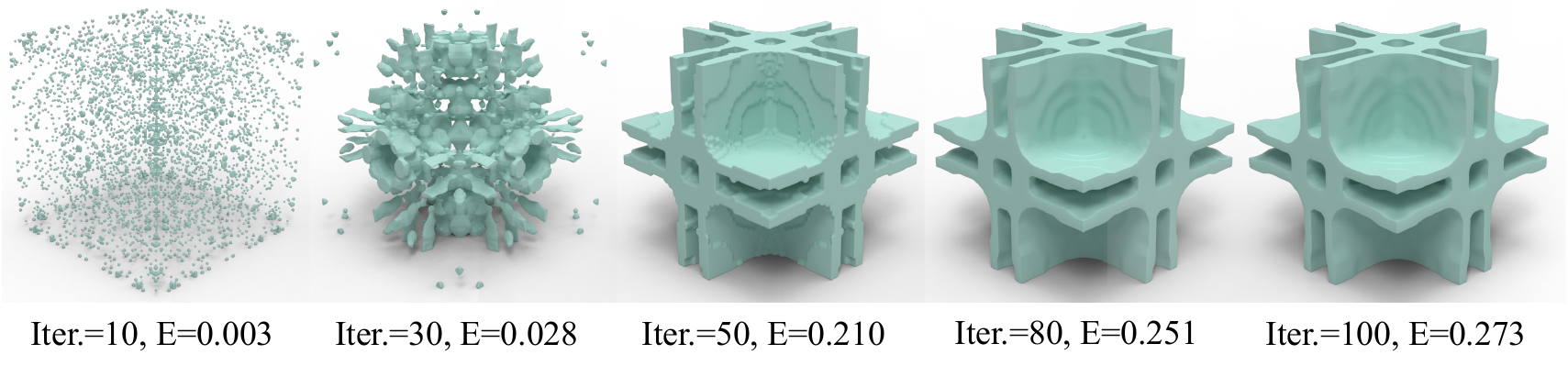}
  \includegraphics[width=\linewidth]{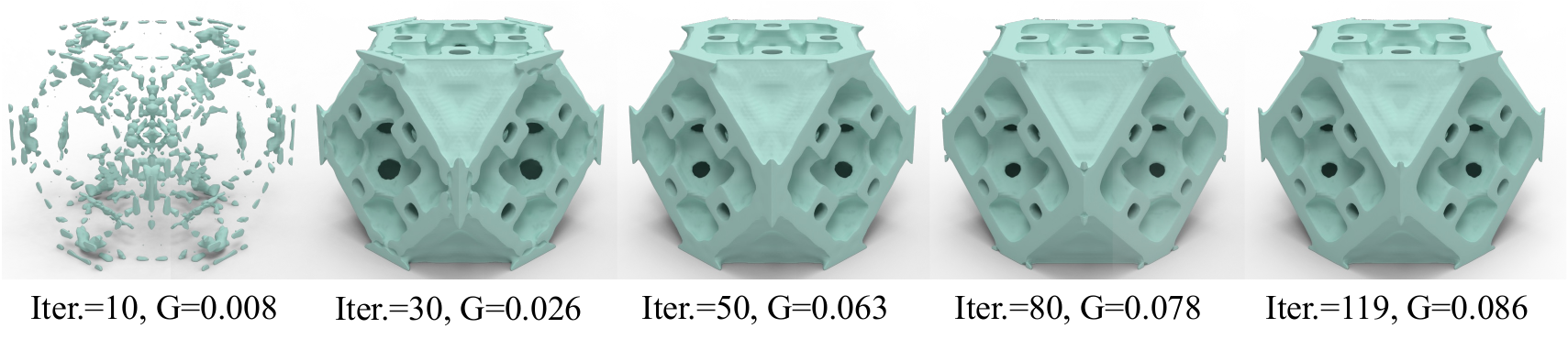}
  \includegraphics[width=\linewidth]{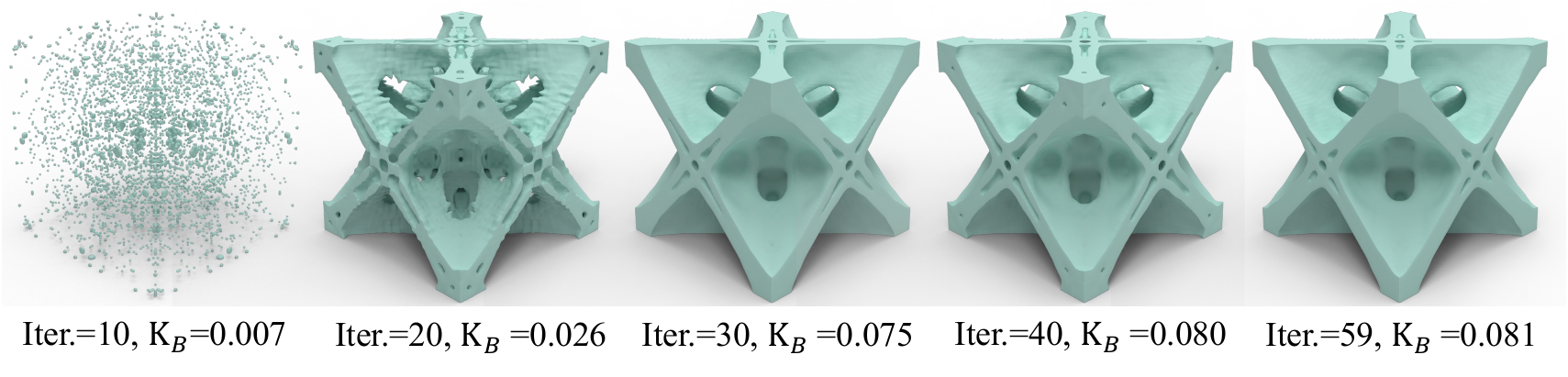}
  \caption{\label{fig:TO-model}
    \textbf{Lattice topology optimization via inverse homogenization.}
    Topology optimization trajectories targeting Young's modulus $E$ (top), shear modulus $G$ (middle), and bulk modulus $K_B$ (bottom), with intermediate snapshots shown along the optimization path.
  }
\end{figure}

%% file: figures/TO-expand.tex
\begin{figure}[t]
  \centering
  \includegraphics[width=\linewidth]{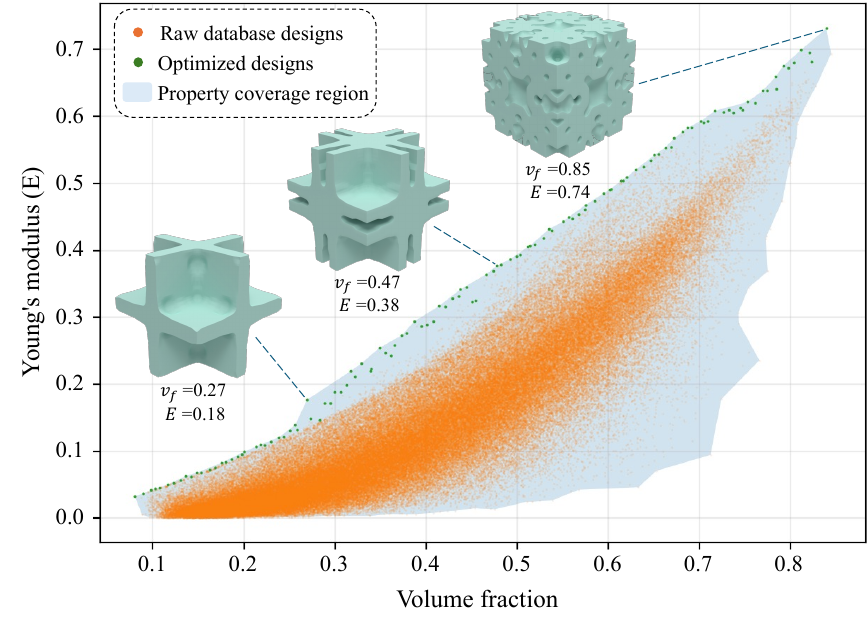}
  \includegraphics[width=\linewidth]{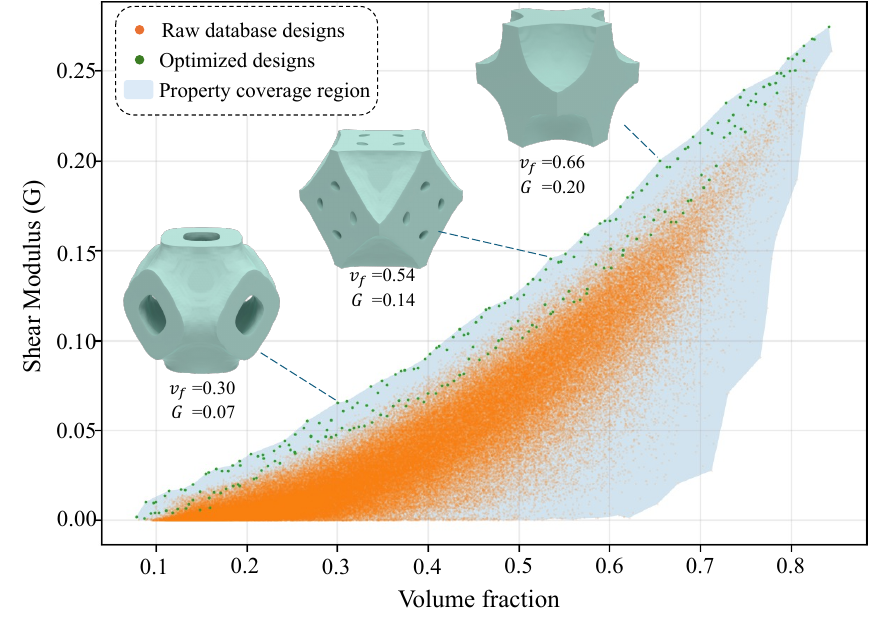}

  \caption{\label{fig:TO-expandEG}
    \textbf{Property-space expansion via inverse homogenization.}
    Starting from extreme-property unit cells in the original database, gradient-based topology optimization is used to maximize Young's modulus $E$ (top) and shear modulus $G$ (bottom).
  }
\end{figure}

%% file: figures/pareto.tex
\begin{figure}[t]
  \centering
\includegraphics[width=.9\linewidth]{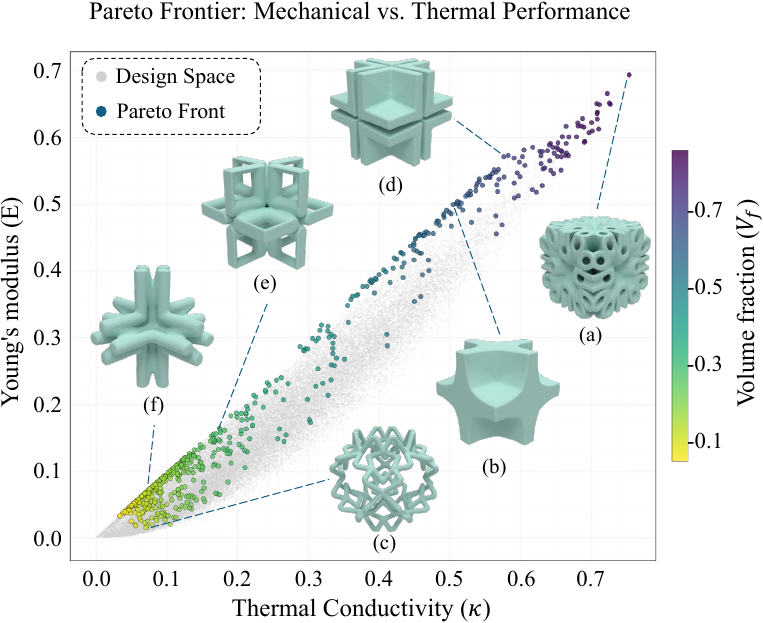}
\caption{\label{fig:Pareto}\textbf{Multi-objective Pareto front analysis.} (a)-(f) showcase representative structures selected for diverse engineering priorities, ranging from absolute performance limits to multi-field synergy under specific density constraints.}
\end{figure}

%% file: src/8-discussion.tex
\section{Conclusion}
\label{sec:discussion}

We have presented \name, a Geometric Multigrid Transformer that bridges the gap between the efficiency of data-driven surrogates and the rigor of numerical solvers for microstructure homogenization. By introducing the principle of architectural alignment, we re-architect the neural backbone to be structurally isomorphic to the Geometric Multigrid hierarchy. This design enables the network to internalize the multi-scale error correction logic, thereby overcoming the spectral bias of pure neural solvers.

Our results demonstrate that \name achieves engineering-grade accuracy ($10^{-5}$ relative residuals) with orders-of-magnitude speedups over state-of-the-art GPU solvers, all while ensuring physical consistency via our  resolution-aware rotary positional encoding (Ra-RoPE). Beyond acceleration, \name significantly lowers the barrier to high-fidelity simulation, enabling real-time ``Generate-and-Filter'' pipelines and interactive inverse design that were previously computationally prohibitive.

\paragraph{Limitations and Future Work}
While \name points toward a new paradigm, our current implementation focuses on periodic homogenization.
Two key challenges remain to extend this framework into a general-purpose solver.
From Homogenization to General FEM: Currently, \name relies on PBCs and fixed canonical loading modes intrinsic to homogenization. Extending the solver to handle arbitrary boundary conditions (\eg Neumann) and flexible force vectors is essential for evolving \name into a general FEM solver capable of broader structural analysis.
Scalability via Distributed Computing: While $512^3$ resolution is sufficient for detailed microstructures, single-GPU memory constraints limit applications requiring billion-degree-of-freedom meshes. Future work will incorporate model parallelism and distributed computing strategies to scale \name beyond current hardware limits.
We will further explore the deep fusion of neural and numerical methods, such as utilizing \name as a self-adjoint preconditioner for the Conjugate Gradient method~\cite{wu2015system, liu2018narrow}, thereby further enhancing convergence robustness for extremely high-contrast microstructures.

Looking forward, \name suggests a broader paradigm of Neural-Numerical Duality, underpinned by Architectural Alignment.
Strictly aligning neural architectures with classical numerical hierarchies allows attention mechanisms to function as physics-aware operators with well-defined numerical roles.
This structural isomorphism provides a rigorous foundation for future differentiable solvers that are data-efficient and physically robust, reshaping, in part, how we design and simulate complex material systems.

%% file: src/appendix.tex
\section{Network Configurations} \label{appendix:A}
Our network adopts a U-Net-like architecture to efficiently map microstructures to physical fields. The architecture comprises three main stages: an encoder, an attention module, and a decoder. The encoder utilizes three layers of sparse convolutions (SpConv~\cite{spconv2022}) to extract local geometric features from the input voxel grid. These encoded features are subsequently fed into the attention mechanism to capture long-range global dependencies. Finally, the processed representations are passed to the decoder to reconstruct the high-resolution dense physical field (\eg  $\hat{\mathbf{u}}^{1}$ or $\{\hat{\mathbf{e}}^{l}\}_{l=2}^{L}$).

\subsection{Base Attention Module Configurations} \label{appendix:A1}

To balance computational efficiency with the ability to capture multi-scale features, we employed a 5-level Geometric Multigrid (GMG) hierarchy ($L=5$) for standard resolution experiments (\eg from $64^3$ to $256^3$). The detailed hyperparameters for this basic configuration are presented in~\cref{tab:model_config}. The number of feature channels doubles as the hierarchy deepens to capture low-frequency global physical modes, while the number of Transformer blocks remains constant across levels. For the smoother, we adopted an asymmetric Gauss-Seidel iteration strategy, gradually increasing the number of iterations as the grid becomes coarser.
\input{tables/A-network-config}

\subsection{Special Configuration for $512^3$ Resolution} \label{appendix:A2}

In the ultra-high resolution scenario of $512^3$, the number of computational nodes increases cubically, posing significant challenges to GPU memory and computational resources. To address this, we adjusted the network architecture by adopting a depth-over-width strategy, increasing the GMG hierarchy to 6 levels ($L=6$). By adding a coarser 6th level, the model can more effectively eliminate extremely low-frequency error components. Simultaneously, to control peak memory usage, we significantly reduced the parameter density at each layer. The specific configuration for the $512^3$ resolution is as follows:
\begin{enumerate}[label=(\alph*),leftmargin=*]
  \item \textbf{Channels}: Reduced to $[12, 24, 48, 96, 192, 192]$ to minimize memory footprint.
  \item \textbf{Window Sizes}: Set to $[64, 128, 256, 512, 1024, 1024]$ to align with the refined grid scales.
\end{enumerate}

To enable training at $512^3$ resolution, we implemented the following low-level optimizations:
\begin{enumerate}[leftmargin=*]
  \item[-] \textbf{Chunk-based Processing \& Memory Optimization}: For operations such as Attention, RoPE positional encoding, and GMG Pooling/Unpooling,we refactored global tensor operations into chunk-based pipeline execution.
  \item[-] \textbf{Block-level Activation Checkpointing}: We enabled checkpointing for the most memory-intensive Transformer Blocks (particularly at the finest level). This mechanism discards intermediate activations during the forward pass and recomputes them during the backward pass.
  \item[-] \textbf{FlashAttention Adaptation \& Mixed Precision}: We ensured that the dimension of each head aligns with FlashAttention kernel shape requirements and utilized bf16 mixed-precision training throughout.
\end{enumerate}

\subsection{Training Details \& Hyperparameters} \label{appendix:A3}
\paragraph{Training setup}
The network is implemented in PyTorch. We optimize the parameters using the AdamW optimizer with a weight decay of $10^{-2}$. The learning rate $\eta$ follows a cosine annealing schedule with warm restarts (CosineAnnealingWarmRestarts, $T_0=10$), configured with a maximum learning rate of $4\times10^{-4}$. The training objective minimizes the physics-informed loss defined by Eq.~(13) in the main paper. To ensure numerical stability, we apply gradient clipping with a global norm capped at 1.0 and utilize gradient accumulation. Additionally, we maintain an Exponential Moving Average (EMA) of the model parameters with a decay rate of 0.999, using the EMA weights during validation and inference. We randomly hold out 5\% of the training set for validation and monitor the relative residual metric to save the best-performing checkpoint.

To balance memory constraints with computational throughput, we employ resolution-dependent batch sizes and hardware configurations. Specifically, the $64^3$ and $128^3$ models were trained on 8 NVIDIA RTX 5090 (\SI{32}{GB}) GPUs for \SI{68}{\hour} and \SI{188}{\hour}, respectively, using per-GPU batch sizes of 8 and 2. For high-resolution experiments, the $256^3$ and $512^3$ models were trained on 2 NVIDIA A100 (\SI{80}{GB}) GPUs for \SI{74}{\hour} and \SI{203}{\hour}, with batch sizes set to 4 and 1. All ablation variants were trained from scratch to ensure consistent and rigorous comparisons. Furthermore, elasticity and conductivity models were trained separately.

\paragraph{Training for Inverse Homogenization}
For inverse-homogenization tasks, the discrete binary occupancy field is relaxed to a continuous density field $\rho \in [10^{-5}, 1]$ following the standard Solid Isotropic Material with Penalization (SIMP) formulation. The model is re-trained using the re-processed and augmented datasets detailed in~\cref{appendix:C}, which incorporate stochastic density values and synthetic periodic structures to enhance numerical robustness during high-noise initializations. All other training configurations, including the AdamW optimizer parameters, learning rate schedule, and EMA decay, remain identical to the standard homogenization training protocol described above. Furthermore, the overall solver pipeline remains completely unchanged, consisting of a single neural forward pass followed by exactly one GMG V-cycle as shown in main text Fig. 2. Importantly, the same re-trained GMT model can be used across different inverse-homogenization objectives, such as maximizing $E$, $G$, or $K$, because GMT replaces the density-aware homogenization solver itself rather than being specialized to a particular optimization objective.

\section{Runtime Decomposition} \label{appendix:B}
\subsection{Baseline Runtime Decomposition} \label{appendix:B1}
\label{app:baseline_time_breakdown}

The wall-clock runtime for numerical baselines (Table~1 in the main paper) is decomposed into:
(i) \emph{initialization time} $T_{\text{init}}$, and
(ii) \emph{solver time} $T_{\text{solve}}$.

$T_{\text{init}}$ includes all one-time setup costs prior to iteration, such as solver context initialization and operator construction. For matrix-based methods, this covers stiffness matrix and load vector assembly; for multigrid methods, hierarchy setup (stencils, restriction/prolongation operators).

$T_{\text{solve}}$ measures the iterative solve only (\eg sparse matrix-vector multiplication (SpMV), smoothing, restriction, prolongation), excluding setup. The total runtime is:
\begin{equation}
  T_{\text{total}} = T_{\text{init}} + T_{\text{solve}}.
\end{equation}

In~\cref{tab:baseline_timing_breakdown}, we report this breakdown for the two fastest GPU baselines, \texttt{AmgX} and \texttt{GMG}, across all three periodic test sets and two representative resolutions per set.

\newcommand{\NA}{\multicolumn{1}{c}{--}}

\begin{table}[t]
  \centering
  \caption{Runtime breakdown of GPU baselines. Timings (in seconds) are averaged over the evaluated test sets.
  }
  \label{tab:baseline_timing_breakdown}
  \scalebox{0.85}{
    \setlength{\tabcolsep}{6pt}
    \begin{tabular}{l
        S[table-format=2.3] S[table-format=2.3] S[table-format=2.3]
        S[table-format=2.3] S[table-format=2.3] S[table-format=2.3]}
      \toprule
                            & \multicolumn{3}{c}{\texttt{AmgX} (GPU)} & \multicolumn{3}{c}{\texttt{GMG} (GPU)}                                    \\
      \(N_{\mathrm{res}}\)  &
      {\(T_{\text{init}}\)} & {\(T_{\text{solve}}\)}                  & {\(T_{\text{total}}\)}                 &
      {\(T_{\text{init}}\)} & {\(T_{\text{solve}}\)}                  & {\(T_{\text{total}}\)}                                                    \\
      \midrule

      64                    & 0.15                                    & 1.48                                   & 1.63   & 0.51  & 2.82   & 3.33   \\
      128                   & 2.31                                    & 12.64                                  & 14.95  & 1.89  & 10.22  & 12.11  \\
      256                   & 8.12                                    & 60.80                                  & 68.92  & 4.03  & 46.61  & 50.64  \\
      512                   & 20.46                                   & 421.06                                 & 441.52 & 29.98 & 359.31 & 389.29 \\

      \bottomrule
    \end{tabular}
  }
  \vspace{-2mm}
\end{table}

\subsection{\name Runtime Decomposition} \label{appendix:B2}
\cref{tab:gmt_timing_breakdown} provides a detailed breakdown of the wall-clock runtime for \name across varying resolutions, ranging from $64^3$ to $512^3$.
The total execution time is decomposed into the specific contributions of the neural network components (Encoder, Attention \& Pooling, Decoder) and the subsequent numerical correction (EBE-GMG).
\begin{enumerate}[label=(\alph*),leftmargin=*]
  \item \textbf{Neural Network Components}: The \texttt{Encoder} and \texttt{Decoder} modules are computationally lightweight; even at the highest resolution of $512^3$, they account for only \SI{0.072}{\second} and \SI{0.112}{\second}, respectively. The \texttt{Attention \& Pooling} module constitutes the majority of the neural inference time (\eg \SI{0.593}{\second} at $512^3$), reflecting the computational cost of capturing long-range global dependencies and performing hierarchical feature aggregation.
  \item \textbf{Numerical Correction (EBE-GMG)}: This column records the time required for the single element-by-element GMG V-cycle that follows the neural initialization. At lower resolutions ($64^3$), this step is very fast (\SI{0.016}{\second}), taking less time than the attention mechanism. At higher resolutions ($512^3$), the numerical solver becomes the dominant component (\SI{1.601}{\second}), yet the total runtime of \SI{2.378}{\second} remains significantly faster than traditional baselines.
\end{enumerate}

\begin{table}[t]
  \centering

  \caption{Runtime breakdown of \name. Timings (in seconds) are averaged over the evaluated test sets.}
  \label{tab:gmt_timing_breakdown}
  \scalebox{0.83}{
    \setlength{\tabcolsep}{6pt}
    \begin{tabular}{l
        S[table-format=2.4] S[table-format=2.4] S[table-format=2.4] S[table-format=2.4] S[table-format=2.4]}
      \toprule
      \(N_{\mathrm{res}}\) & \multicolumn{1}{c}{\texttt{Encoder}} & \multicolumn{1}{c}{\texttt{Attention \& Pooling}} & \multicolumn{1}{c}{\texttt{Decoder}} & \multicolumn{1}{c}{\texttt{EBE-GMG}} & \multicolumn{1}{c}{\texttt{Total}} \\
      \midrule
      64                   & 0.002                                & 0.027                                             & 0.003                                & 0.016                                & 0.048                              \\
      128                  & 0.002                                & 0.053                                             & 0.004                                & 0.064                                & 0.123                              \\
      256                  & 0.013                                & 0.330                                             & 0.015                                & 0.237                                & 0.595                              \\
      512                  & 0.072                                & 0.593                                             & 0.112                                & 1.601                                & 2.378                              \\

      \bottomrule
    \end{tabular}
  }
\end{table}

\section{Dataset Details} \label{appendix:C}

To ensure reproducibility and facilitate rigorous benchmarking, this section details the generation protocols and geometric profiles of the datasets used in this work. Visual examples of these geometric classes are provided in~\cref{fig:dataset-profile}.
\input{figures/A-dataset-profile}
\paragraph{Generation Protocol and Distribution}
The construction of our training datasets (TPMS, PSL, and Truss) follows the established methodologies described in~\cite{xing2025pcg}. For the training set, $V_f$ is uniformly sampled within the range of $[10\%, 30\%]$.
\paragraph{Geometric Profiles}
The dataset consists of five topological categories covering diverse geometric complexity:
\begin{enumerate}[label=(\alph*),leftmargin=*]
  \item \textbf{Triply Periodic Minimal Surfaces (TPMS)}: Generated via level-set functions (\eg Gyroid, Diamond, Primitive), these structures are characterized by smooth, continuous interfaces and high surface-to-volume ratios. They represent geometry-dominated features with high curvature.
  \item \textbf{Truss-like Structures (Truss)}~\cite{panetta2015elastic, zheng2023unifying}: Constructed by connecting spatially distributed nodes with cylindrical beams. These geometries feature discrete strut connectivity and sharp intersections, challenging the solver's ability to handle high-frequency geometric discontinuities.

  \item \textbf{Parametric Shell Lattices (PSL)}~\cite{liu2022parametric}: Generated using a skeleton-based parametric modeling approach. This method defines a geometric skeleton and applies parametric control functions to generate continuous, smooth shell surfaces.
  \item \textbf{Sto-MS (Non-Periodic Test Set)}~\cite{Martvoronoifoams2016}: Comprising procedural, aperiodic microstructures inspired by Voronoi open-cell foams. These geometries feature open boundaries and are used to benchmark the solver's robustness in non-periodic settings, where the problem is numerically treated as a generic linear system.
  \item \textbf{L-BOM (OOD Test Set)}~\cite{wang2026data}: Microstructures generated via data-driven inverse design. This dataset represents complex geometries, testing the solver's generalization to ML-generated topologies.
\end{enumerate}

To quantitatively assess the distributional similarity between the training data and various test sets, we employ the Intersection over Union (IoU) metric to measure the structure similarity between binary occupancy fields. The measured IoU values for the test sets relative to the training set are: TPMS ($0.92$), PSL ($0.77$), Truss ($0.75$), L-BOM ($0.34$), and Sto-MS ($0.26$). The significantly lower IoU scores for L-BOM and Sto-MS confirm that they represent distributions far from the primary training data, thereby validating their use as rigorous out-of-distribution (OOD).

\paragraph{Data Augmentation for Inverse Homogenization}
To support the differentiable topology optimization and inverse design tasks, we re-processed our geometric datasets and re-trained the GMT model specifically on continuous density fields. For the original samples in the TPMS, PSL, and Truss categories, binary occupancy was replaced by random density values sampled from $[10^{-5}, 0.99999]$. This modification allows the network to learn the material interpolation and sensitivity information essential for gradient-based optimization. Furthermore, to enhance the numerical stability of GMT during the high-noise initializations typical of inverse design, we introduced an additional dataset of \num{10000} synthetic random structures. These structures are generated with stochastic density distributions while strictly enforcing periodic boundary conditions. This comprehensive re-training ensures that the model provides robust and differentiable property evaluations across the entire continuous design space.

\section{Additional Experimental Results} \label{appendix:D}
\subsection{Additional stress tests} \label{appendix:D1}

Beyond the main fixed-material setting of the paper, we evaluate GMT on two more challenging cases. First, on random binary microstructures outside the main training families, the pretrained GMT still achieves a relative residual of $8.03 \times 10^{-5}$ without retraining. Second, for heterogeneous/high-contrast materials, we replace binary occupancy inputs with local material features and retrain GMT over $E \in [10^{-5},1]$ and $\nu \in [0.2,0.4]$. Under this material-aware setting, GMT remains stably convergent, with residuals ranging from $3.62 \times 10^{-5}$ to $2.97 \times 10^{-4}$ as the contrast increases. When further evaluated on unseen higher contrasts up to $E \in [10^{-8},1]$, GMT still reaches $5.12 \times 10^{-3}$. These results suggest that GMT generalizes well to irregular binary geometries, while for strong material heterogeneity it can be extended effectively through material-aware inputs and retraining.

\subsection{More accuracy} \label{appendix:D2}

  Higher precision is not a limitation of GMT. In the main paper, the W-cycle variant already reaches a residual of $8.13 \times 10^{-6}$. Moreover, GMT can naturally serve as a warm start for additional numerical refinement: by appending 8 standard V-cycles after the GMT prediction, we reach a residual of $10^{-7}$. To achieve the same precision, Zhang-GMG and AMG require 110 and 130 iterations, respectively. Therefore, GMT is not restricted to the default $10^{-5}$ regime, but also provides an efficient initialization for stricter high-accuracy solves.

\begin{table}[t]
  \centering
  \caption{
    Pure compute-time comparison (measured in seconds). We isolate the core computation time of each method rather than end-to-end latency. For the numerical baselines, we report only the V-cycle solve time, excluding initialization, hierarchy/operator construction, and assembly costs. For GMT, we report only a single inference pass, consisting of the network forward pass followed by one GMG forward refinement, excluding data loading and feature assembly. Speedup is computed against this pure GMT inference time. }
  \label{tab:solve_only_time}
  \scalebox{0.85}{
    \begin{tabular}{c|r|rr|rr}
      \toprule
                                                                 &                & \multicolumn{2}{c|}{\texttt{AmgX}} & \multicolumn{2}{c}{\texttt{GMG}}                            \\
      \cmidrule(lr){3-4}\cmidrule(lr){5-6}
      $\mathbf{N_{res}}$                                         &
      \multicolumn{1}{c|}{\(T^{\texttt{GMT}}_{\text{Forward}}\)} &
      \multicolumn{1}{c}{\(T_{\text{solve}}^{\texttt{AmgX}}\)}   &
      \multicolumn{1}{c|}{Speedup}                               &
      \multicolumn{1}{c}{\(T_{\text{solve}}^{\texttt{GMG}}\)}    &
      \multicolumn{1}{c}{Speedup}                                                                                                                                                    \\
      \midrule
      64                                                         & \textbf{0.046} & 1.48                               & 32.2$\,\times$                   & 2.82   & 61.3$\,\times$  \\
      128                                                        & \textbf{0.119} & 12.64                              & 106.2$\,\times$                  & 10.22  & 85.9$\,\times$  \\
      256                                                        & \textbf{0.589} & 60.80                              & 103.2$\,\times$                  & 46.61  & 79.1$\,\times$  \\
      512                                                        & \textbf{2.360} & 421.06                             & 178.4$\,\times$                  & 359.31 & 152.3$\,\times$ \\
      \bottomrule
    \end{tabular}
  }
\end{table}

\subsection{Pure Compute-Time Comparison Excluding Setup Overheads} \label{appendix:D3}
To further address the concern about $T_{init}$ amortization, we additionally report a pure compute-time comparison in Table~\ref{tab:solve_only_time}. In this experiment, we remove non-solver overheads from both sides and compare only the core computation. Specifically, for the numerical baselines, we retain only the V-cycle runtime and discard initialization, hierarchy/operator construction, and assembly costs. For GMT, we retain only a single inference pass, consisting of the network forward pass and one GMG forward refinement, while excluding data loading and feature assembly. Therefore, this comparison focuses on solver-core computation rather than end-to-end latency. Even under this setup, which removes substantial non-recurring overhead from the numerical methods, GMT remains consistently faster across all tested resolutions, showing that the advantage of GMT is not merely due to setup costs, but also persists in the pure computation stage itself.

\section{Matrix-Free Geometric Transfer Operators} \label{appendix:E}

Effective coarse-grid correction requires accurate transfer of residuals and error corrections between hierarchy levels. We employ topology-consistent trilinear transfer operators implemented in a strictly matrix-free manner to minimize memory footprint and maximize parallelism.

\subsection{Geometric Mapping and Weight Calculation} \label{appendix:E1}
Consider a fine grid node at coordinate $\mathbf{x}_f$ and its corresponding parent coarse cell index $\mathbf{x}_c = \lfloor \mathbf{x}_f / 2 \rfloor$. The coarse neighborhood is defined by the eight corner offsets $o_k \in \{0,1\}^3$. Under periodic boundary conditions, the coordinate of the $k$-th coarse neighbor $\mathbf{x}_c^{(k)}$ is given by:
\begin{equation}
  \mathbf{x}_c^{(k)} = (\mathbf{x}_c + o_k) \pmod{N_{c}},
\end{equation}
where $N_{c}$ denotes the coarse grid resolution.

To determine the interpolation weights, let $\rho = \mathbf{x}_f \pmod 2 \in \{0,1\}^3$ represent the local integer offset of the fine node within the coarse cell. We define the normalized local coordinate $\boldsymbol{\xi} = \rho / 2 \in \{0, 0.5\}^3$. The trilinear interpolation weight $w_k$ for the $k$-th coarse neighbor is then computed as the product of 1D linear shape functions:
\begin{equation}
  w_k(\mathbf{x}_f) = \prod_{d \in \{x,y,z\}} \big( 1 - |\xi_d - o_{k,d}| \big).
\end{equation}

\subsection{Matrix-Free Implementation} \label{appendix:E2}
To enable efficient GPU execution, we avoid explicit matrix assembly. Instead, we precompute a lightweight stencil tuple $(\mathcal{I}, \mathbf{W})$ for each active fine node $i$. Here, $\mathcal{I}_{i,k}$ stores the flattened memory index of the $k$-th coarse neighbor (accounting for periodicity), and $\mathbf{W}_{i,k} = w_k(\mathbf{x}_{f,i})$ stores the associated weight.

\textbf{Prolongation (Coarse-to-Fine):} The prolongation operator $\mathbf{P}$ interpolates corrections from the coarse grid to the fine grid. This is implemented as a parallel \textit{weighted gather} operation:
\begin{equation}
  \mathbf{u}_f^{(\ell)}(i) = \sum_{k=1}^8 \mathbf{W}_{i,k} \cdot \mathbf{u}_c^{(\ell+1)}\big( \mathcal{I}_{i,k} \big).
\end{equation}

\textbf{Restriction (Fine-to-Coarse):} To ensure energy consistency, the restriction operator is strictly defined as the transpose of prolongation ($\mathbf{R} = \mathbf{P}^T$). In our matrix-free framework, this is realized via an atomic \textit{scatter-add} operation, where fine-grid residuals are accumulated onto coarse nodes:
\begin{equation}
  \mathbf{r}_c^{(\ell+1)}(j) \leftarrow \sum_{(i,k) : \mathcal{I}_{i,k}=j} \mathbf{W}_{i,k} \cdot \mathbf{r}_f^{(\ell)}(i).
\end{equation}
This formulation ensures operator symmetry and consistent periodic boundary handling without global matrix storage.

\section{Homogenization Theory} \label{appendix:F}
This section presents the finite element formulation for homogenization analysis of periodic lattice structures as implemented in the \name framework~\cite{andreassen2014determine, dong2019149}. The Representative Volume Element (RVE) is denoted by $\Omega$. Two physical domains are considered: linear elasticity (mechanical) and steady-state heat conduction (thermal).

\subsection{Linear Elasticity (Mechanical)} \label{appendix:F1}

The homogenized elastic properties are represented by the effective stiffness tensor $\mathbf{C}^H \in \mathbb{R}^{6 \times 6}$ (Voigt notation). The goal is to compute the relationship between macroscopic strain $\bar{\boldsymbol{\epsilon}}$ and the volume-averaged microscopic stress~\cite{dong2019149}.

\paragraph{Governing Equations}

For a prescribed macroscopic strain $\bar{\boldsymbol{\epsilon}}$, the displacement corrector $\mathbf{u} \in \mathbb{R}^3$ satisfies:
\begin{equation}
  \nabla \cdot \big( \mathbf{C}_0 : (\boldsymbol{\epsilon}(\mathbf{u}) + \bar{\boldsymbol{\epsilon}}) \big) = 0 \quad \text{in } \Omega,
\end{equation}
where $\mathbf{C}_0$ is the base elasticity tensor. We solve this for six independent unit strain modes:
\[
  \bar{\boldsymbol{\epsilon}}_{11}=(1,0,0,0,0,0)^\mathsf{T}, \;
  \bar{\boldsymbol{\epsilon}}_{22}=(0,1,0,0,0,0)^\mathsf{T}, \;
  \bar{\boldsymbol{\epsilon}}_{33}=(0,0,1,0,0,0)^\mathsf{T},
\]
\[
  \bar{\boldsymbol{\epsilon}}_{23}=(0,0,0,1,0,0)^\mathsf{T}, \;
  \bar{\boldsymbol{\epsilon}}_{13}=(0,0,0,0,1,0)^\mathsf{T}, \;
  \bar{\boldsymbol{\epsilon}}_{12}=(0,0,0,0,0,1)^\mathsf{T}.
\]

\paragraph{Weak Formulation and Discretization}

The weak Galerkin formulation  for the displacement field $u$ is:
\begin{equation}
  \int_{\Omega} \boldsymbol{\epsilon}(\mathbf{v}) : \mathbf{C}_0 : \boldsymbol{\epsilon}(\mathbf{u}) \, d\Omega
  = - \int_{\Omega} \boldsymbol{\epsilon}(\mathbf{v}) : \mathbf{C}_0 : \bar{\boldsymbol{\epsilon}} \, d\Omega, \quad \forall \mathbf{v}.
\end{equation}

The RVE is discretized into $N_{\text{res}}^3$ eight-node hexahedral elements. For each element $e$:
\[
  \mathbf{u}_e = \mathbf{N} \mathbf{d}_e, \quad \boldsymbol{\epsilon}(\mathbf{u})_e = \mathbf{B} \mathbf{d}_e,
\]
where $\mathbf{N}$ is the shape function matrix and $\mathbf{B}$ is the strain-displacement matrix. The element stiffness matrix $\mathbf{K}_e \in \mathbb{R}^{24 \times 24}$ and load vector $\mathbf{f}_e \in \mathbb{R}^{24 \times 6}$ are:
\[
  \mathbf{K}_e = \int_{\Omega_e} \mathbf{B}^\mathsf{T} \mathbf{C}_0 \mathbf{B} \, d\Omega, \quad
  \mathbf{f}_e = \int_{\Omega_e} \mathbf{B}^\mathsf{T} \mathbf{C}_0 \bar{\boldsymbol{\epsilon}} \, d\Omega.
\]

\paragraph{Effective Property Calculation}
After solving $\mathbf{K}\mathbf{u} = \mathbf{f}$, the homogenized tensor $\mathbf{C}^H$ is computed by summing the contributions of all active voxels:
\begin{equation}
  C_{ijkl}^H = \frac{1}{|\Omega|} \sum_{e \in \Omega} (\mathbf{x}_0 - \mathbf{u}_e)^\mathsf{T} \mathbf{K}_e (\mathbf{x}_0 - \mathbf{u}_e),
\end{equation}
where $\mathbf{x}_0 = \mathbf{K}_e^{-1}\mathbf{f}_e$ is the displacement under uniform strain without periodic correction.

\subsection{Steady Heat Conduction (Thermal)} \label{appendix:F2}

The homogenized thermal conductivity $\boldsymbol{\kappa}^H \in \mathbb{R}^{3 \times 3}$ relates the macroscopic temperature gradient $\bar{\mathbf{g}}$ to the averaged heat flux~\cite{QUAN2026100118}.

\paragraph{Governing Equation}
For each loading mode $i \in \{x,y,z\}$, the temperature corrector $\psi_i$ satisfies:
\begin{equation}
  \nabla \cdot  \kappa_0(\nabla \psi_i + \mathbf{e}_i)  = 0 \quad \text{in } \Omega,
\end{equation}
where $\kappa_0$ is the base heat conductivity and $\mathbf{e}_i$ is the unit vector along axis $i$.

\paragraph{Weak Formulation and Discretization}
For each loading mode $i\in\{x,y,z\}$, the weak form is: find $\psi_i$ such that for all admissible $v$,
\begin{equation}
  \int_{\Omega} \nabla v \cdot \kappa_0 \nabla \psi_i \, d\Omega
  = -\int_{\Omega} \nabla v \cdot \kappa_0 \mathbf{e}_i \, d\Omega.
\end{equation}

Using the same voxel discretization, the element conductivity matrix
$\mathbf{K}_e^{\text{th}} \in \mathbb{R}^{8 \times 8}$ and the element load matrix
$\mathbf{F}_e^{\text{th}} \in \mathbb{R}^{8 \times 3}$ (three loading modes assembled at once) are
\[
  \mathbf{K}_e^{\text{th}} = \int_{\Omega_e} \mathbf{B}_{\text{th}}^\mathsf{T} \kappa_0 \mathbf{B}_{\text{th}} \, d\Omega, \qquad
  \mathbf{F}_e^{\text{th}} = \int_{\Omega_e} \mathbf{B}_{\text{th}}^\mathsf{T} \kappa_0 \mathbf{I}_{3\times 3} \, d\Omega,
\]
where $\mathbf{B}_{\text{th}}$ maps nodal temperatures to gradients, and
$\mathbf{I}_{3\times 3}=[\mathbf{e}_x,\mathbf{e}_y,\mathbf{e}_z]$.

Assembling all elements yields the global linear system
\begin{equation}
  \mathbf{K}^{\text{th}} \boldsymbol{\psi}^{(i)} = \mathbf{F}^{\text{th}{(i)}}.
\end{equation}

\paragraph{Effective Property Calculation}
Let $T_0^{(i)}(\mathbf{x})=\mathbf{e}_i\cdot\mathbf{x}$ be the affine temperature field producing a unit macroscopic
gradient in direction $i$, and let $\mathbf{T}_{0,e}^{(i)}$ denote its nodal vector on element $e$.

The homogenized conductivity tensor is computed as
\begin{equation}
  \kappa_{ij}^H = \frac{1}{|\Omega|} \sum_{e \in \Omega}
  \big(\mathbf{T}_{0,e}^{(i)} - \boldsymbol{\psi}_e^{(i)}\big)^\mathsf{T}\,
  \mathbf{K}_e^{\text{th}}\,
  \big(\mathbf{T}_{0,e}^{(j)} - \boldsymbol{\psi}_e^{(j)}\big).
\end{equation}

\subsection{Lattice topology optimization via inverse homogenization} \label{appendix:F3}
\paragraph{Inverse Homogenization Setup.}
We perform unit-cell topology optimization on the periodic RVE $\Omega$, discretized into $N_{\mathrm{res}}^3$ voxel elements.
The design variable is the elementwise density field $\boldsymbol{\rho}=\{\rho_e\}_{e=1}^{N_{\mathrm{res}}^3}$ with
\(
\rho_{\min} \le \rho_e \le 1,
\)
where $\rho_{\min} > 0$ prevents singular stiffness.
A target volume fraction constraint is imposed:
\begin{equation} \label{eq:constraint}
  \frac{1}{|\Omega|}\sum_{e=1}^{N_{\mathrm{res}}^3} v_e \rho_e \le v_f,
\end{equation}
with $v_e$ the element volume and $v_f$ the prescribed material fraction.

Material interpolation follows a SIMP-type model:
\begin{equation}\label{eq:ch}
  \mathbf{C}_e(\rho_e) = \bigl(\rho_{\min} + \rho_e^{\,p}\bigr)\,\mathbf{C}_0,
\end{equation}
where $\mathbf{C}_0$ is the base elasticity tensor and $p > 1$ is the penalization factor.

\paragraph{Inverse Homogenization Algorithm}
At each iteration of the optimization process, the homogenized elasticity tensor $\mathbf{C}_H(\boldsymbol{\rho}) \in \mathbb{R}^{6 \times 6}$ is obtained by solving the six periodic cell problems corresponding to the canonical unit strain modes~\cite{SIGMUND19942313, groen2018homogenization, sha2022topology}.
Let $\boldsymbol{\chi}^{(m)}$ denote the displacement corrector for mode $m \in \{1,\ldots,6\}$, and let the discretized cell problems be:
\begin{equation}
  \mathbf{K}(\boldsymbol{\rho})\,\mathbf{u}^{(m)} = \mathbf{f}^{(m)}, \qquad m = 1,\ldots,6,
\end{equation}
where $\mathbf{K}(\boldsymbol{\rho})$ is assembled (matrix-free) from the interpolated element tensors $\mathbf{C}_e(\rho_e)$, and $\mathbf{u}^{(m)}$ stores the nodal degrees of freedom.

We solve the six systems using the differentiable EBE–GMG pipeline: \name provides multi-level initial guesses, followed by a fixed number of GMG V-cycles to achieve the required residual accuracy. Since the full pipeline is differentiable, gradients can be obtained by reverse-mode AD.
For efficiency, analytical sensitivities are also supported:
\begin{align}
  \frac{\partial J}{\partial \rho_e}
   & = \sum_{i,j}
  \frac{\partial J}{\partial (\mathbf{C}_H)_{ij}}\,
  \frac{\partial (\mathbf{C}_H)_{ij}}{\partial \rho_e}, \\
  \frac{\partial (\mathbf{C}_H)_{ij}}{\partial \rho_e}
   & = \frac{p\,\rho_e^{\,p-1}}{|\Omega|}
  \bigl(\mathbf{d}^{(i)}_e\bigr)^{\!\top}
  \mathbf{K}_0\,
  \bigl(\mathbf{d}^{(j)}_e\bigr),
\end{align}
where $\mathbf{K}_0$ is the element stiffness matrix associated with $\mathbf{C}_0$, and $\mathbf{d}^{(i)}_e$ is the element-level displacement difference for mode $i$ (uniform-strain component minus periodic correction).

To ensure manufacturability, a standard sensitivity filter is applied to impose a minimum feature size. The density field $\boldsymbol{\rho}$ is updated using a first-order constrained optimizer (Optimality Criteria method with move limits) and iterated until the convergence of either the objective function or the density distribution.

\paragraph{Optimization Targets}
To generate the distinct metamaterial classes shown in Fig. 10 in the main text, we formulate specific objective functions based on the homogenized compliance matrix $\mathbf{S}^H = (\mathbf{C}^H)^{-1}$. We maximize the target stiffness moduli by minimizing their corresponding compliance components:

\begin{enumerate}[label=(\alph*),leftmargin=*]
  \item \textbf{Young's Modulus Optimization}: We maximize the stiffness by minimizing the mean of the principal normal compliance terms ($S_{00}, S_{11}, S_{22}$).
  \item \textbf{Shear Modulus Optimization}: We maximize the shear stiffness by minimizing the mean of the shear compliance terms ($S_{33}, S_{44}, S_{55}$).
  \item \textbf{Bulk Modulus Optimization}: We maximize the bulk modulus by minimizing the volumetric compliance, defined as the sum of the upper-left $3\times3$ block of the compliance matrix ($\sum_{i=0}^{2}\sum_{j=0}^{2} S_{ij}$).
\end{enumerate}

%% file: tables/A-network-config.tex
\begin{table}[t]
    \centering
    \caption{Basic Attention Block Configuration of GMT ($L=5$). The channel dimensions double with each coarser level to capture global features, while the number of Transformer blocks remains constant.}
    \label{tab:model_config}
    \scalebox{0.68}{
        \begin{tabular}{ccc}
            \toprule
            \textbf{Parameter}                            & \textbf{Values}                & \textbf{Description}              \\
            \midrule
            Grid Levels ($L$)                             & 5                              & Number of \name levels            \\
            Channels                                      & $[24, 48, 96, 192, 384]$       & Feature dimensions per level      \\
            Blocks per Level                              & $[4, 4, 4, 4, 4]$              & Number of Transformer blocks      \\
            Attention Heads                               & $[2, 4, 4, 8, 8]$              & Heads in Multi-Head Attention     \\
            Window Sizes                                  & $[256, 512, 1024, 1024, 1024]$ & Serialized attention window sizes \\
            Training Smoothing ($\{\mathbf{It}^l\}_1^L$)  & $\{1, 2, 2, 2, 2\}$            & GS Smoothing iterations           \\
            Inference Smoothing ($\{\mathbf{It}^l\}_1^L$) & $\{2, 2, 2, 2, 2\}$            & GS Smoothing iterations           \\
            \bottomrule
        \end{tabular}
    }
\end{table}

%% file: figures/A-dataset-profile.tex
\begin{figure}[t]
  \centering
\includegraphics[width=\linewidth]{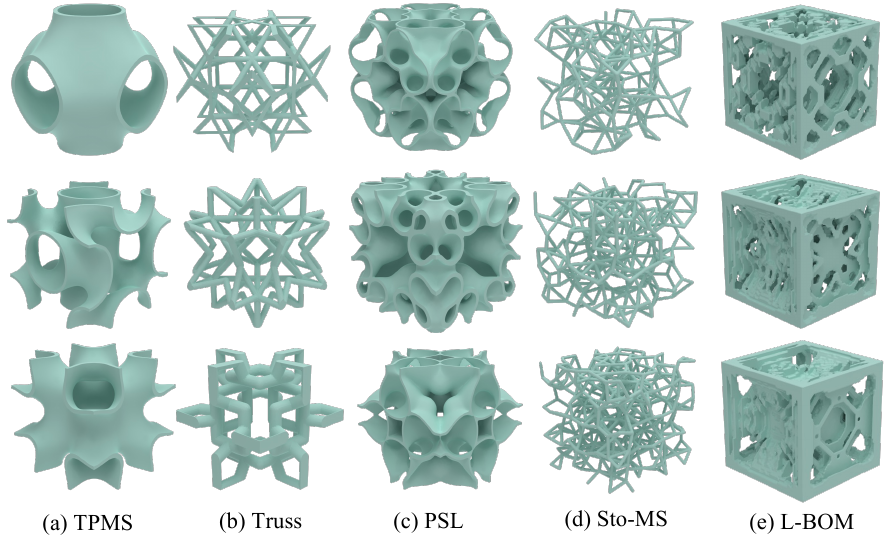}
\caption{\label{fig:dataset-profile}
Visualization of the geometric datasets used in this work:
(a) TPMS, (b) Truss, and (c) PSL (training and standard evaluation);
(d) Sto-MS (non-periodic test set);
and (e) L-BOM (out-of-distribution test set).
}
\end{figure}